\documentstyle[12pt,psfig]{article}
\newcommand{\mysection}{\setcounter{equation}{0}\section}

\def\beq{\begin{equation}}
\def\eeq{\end{equation}}
\def\beqa{\begin{eqnarray}}
\def\eeqa{\end{eqnarray}}

\newlength{\dinwidth} \newlength{\dinmargin}
\begin{document}
\begin {flushright}
FSU-HEP-20000720\\
\end {flushright} 
\vspace{3mm}
\begin{center}
{\Large \bf Effects of Higher-Order Threshold Corrections in 
High-$E_T$ Jet Production}
\end{center}
\vspace{2mm}
\begin{center}
{\large Nikolaos Kidonakis and J. F. Owens}\\
\vspace{2mm}
{\it Physics Department\\
Florida State University\\
Tallahassee, FL 32306-4350, USA} \\
\end{center}

\begin{abstract}
Results for higher-order threshold enhancements in high-$E_T$ jet 
production in hadron-hadron collisions are presented. 
Expressions are given for the next-to-next-to-leading order (NNLO) threshold 
corrections   
to the single-jet inclusive cross section at next-to-leading logarithmic 
(NLL) accuracy. 
The corrections are found to be small for the specific choice of $E_T/2$ for 
the factorization and renormalization scales, and the corrected cross section 
shows a substantial reduction of the scale dependence. 
A comparison to experimental results from the Tevatron is presented.

\end{abstract}
\pagebreak

\mysection{Introduction}

QCD provides an impressive description of the data for the production of 
high-$E_T$ jets in hadronic collisions. The data taken by the CDF \cite  
{CDF} and D\O\  \cite{Abbott:1999ya} Collaborations for the single jet 
inclusive cross 
section fall by nine orders of magnitude in the $E_T$ range extending to 460
GeV, a behavior also seen in the theoretical predictions. Nevertheless, there 
is a hint that the theory slightly underestimates the data towards the upper
end of the measured $E_T$ range. Although such an excess of experiment over
theory could be a sign of new physics, other more conventional explanations are
possible. On the one hand, it could be that the parton distributions at the
large values of $x$ needed for the theoretical calculations are slightly
underestimated. In particular, a somewhat harder gluon distribution than is
usually used might be sufficient to describe the data. This approach has been
studied by the CTEQ Collaboration \cite{Huston:1996tw} 
which developed a special parton set, CTEQ5HJ \cite{Lai:2000wy}, 
to describe these data. On the other hand, it may be that there
are corrections to the hard scattering parton-parton cross sections that have
yet to be included in conventional next-to-leading-order calculations. It is
this latter possibility which is examined in this paper.

At large values of $x_T=2E_T/\sqrt{S}$, 
the phase space for gluon emission is limited by the
rapid decrease of the parton distributions. 
The parton-parton scattering processes are restricted to the threshold region 
where there are large logarithmic corrections.
Threshold resummation is a formalism for including the effects of these
corrections to all orders in perturbation theory. The effects due to threshold
resummation are expected to grow as $x_T$ approaches one. The high-$E_T$ jet
data extend in $x_T$ to about 0.5 and it is of interest to see if some of the 
observed excess of data over theory can be accounted for by the inclusion of 
threshold corrections.

In the next section, the formalism for threshold resummation to NLL
accuracy for single jet
production in hadron-hadron collisions is reviewed. Sec. 3 contains a
comparison of our results with Tevatron data while Sec. 4 contains our
conclusions. Detailed expressions for the NNLO threshold corrections, derived 
from the expansion of the resummed cross section, are
given in a set of appendices for each of the relevant parton-parton
subrocesses.

\mysection{The single-jet inclusive resummed cross section}

In this section we discuss the single-jet inclusive cross section
at large transverse momentum, paying particular attention to
threshold corrections that occur at each order in perturbation theory. 
Consider jet production in $p {\bar p}$ collisions,
\beq
p+{\bar p} \rightarrow J +X \, ,
\eeq
the invariant single-jet inclusive cross section 
for which may be written in factorized form as
\beqa
E_J\frac{d^3\sigma_{p {\bar p}\rightarrow J X}}{d^3 p_J}
&=&\sum_{f} \int dx_a dx_b \, \phi_{f_a/p}(x_a,\mu_F^2) \,
\phi_{f_b/{\bar p}}(x_b,\mu_F^2)
\nonumber \\ &&  \times \, 
E_J \frac{d^3{\hat \sigma}_{f_a f_b \rightarrow J X}}
{d^3  p_J}(s,t,u,\mu_F,\alpha_s(\mu_R^2)) \, .
\label{factjet}
\eeqa
The initial-state collinear singularities have been absorbed into the
parton distribution functions
$\phi$ at a factorization scale $\mu_F$, while $\mu_R$ is the
renormalization scale.
The parton-parton hard scattering cross section is denoted by 
${\hat \sigma}$.

The parton-parton scattering  subprocesses contributing to jet  
production in lowest order are
\beqa
q_j{\bar q_j} & \rightarrow & q_j {\bar q_j} \, , \quad
q_j{\bar q_j}  \rightarrow  q_k {\bar q_k} \, , \quad
q_j{\bar q_k}  \rightarrow  q_j {\bar q_k} \, , \quad
q_j q_j  \rightarrow  q_j q_j \, , \quad
q_j q_k  \rightarrow  q_j q_k \, ,
\nonumber \\
q {\bar q} & \rightarrow & g g \, , \quad
g g  \rightarrow  q {\bar q} \, , \quad
q g  \rightarrow  q g \, , \quad
g g  \rightarrow   g g \, .
\eeqa
For the process $f_a(p_a)+f_b(p_b) \rightarrow J(p_J) + X$, the
Mandelstam invariants constructed from the parton and 
jet four-vectors are given by
\beq
s=(p_a+p_b)^2 \, , \quad t=(p_a-p_J)^2 \, , 
\quad u=(p_b-p_J)^2 \, ,
\eeq
which satisfy $s_4 \equiv s+t+u=0$ at threshold.
The variable $s_4$ is the square of the invariant mass of the system
recoiling against the observed jet. 

The threshold for the above partonic subprocesses occurs at 
$s_4=0$. The values of $x_a$ and $x_b$ corresponding to this point are the 
minimum values which can give rise to a jet of the specified rapidity and 
transverse momentum. Near threshold the phase space for the emission of 
real gluons is limited and large logarithmic corrections arise from the
incomplete cancellation of infrared divergences against 
virtual gluon emission contributions. In general, $\hat{\sigma}$ 
includes plus distributions with respect
to $s_4$ at $n$th order in $\alpha_s$ of the type
\beq
\left[\frac{\ln^{m}(s_4/p_T^2)}{s_4} \right]_+, \hspace{10mm} m\le 2n-1\, ,
\label{s4plus}
\eeq
where $p_T=(tu/s)^{1/2}$ is the transverse momentum of the jet.
These distributions have been resummed to all orders at 
next-to-leading logarithmic (NLL) accuracy 
for dijet and singlet-jet production in Refs.~\cite{Kidonakis:1998bk,
Kidonakis:1998nf,Laenen:1998qw}.

The resummation is achieved in moment space 
through a refactorization \cite{Kidonakis:1998bk,Laenen:1998qw,
Kidonakis:1996aq,Kidonakis:1997gm} of the cross section into a product
of functions $\psi$ and $J$ that absorb the collinear singularities
in the incoming partons and outgoing jets, respectively; a soft gluon function
$S$ that encompasses noncollinear soft gluon emission; and a hard-scattering
function $H$ that describes the short-distance hard-scattering.
The color structure of the hard scattering depends on the flavor content
(for example singlet or octet for $q{\bar q}\rightarrow q{\bar q}$) and
is described by a set of color tensors $c_I$.
Note that the functions $S$ and $H$ are actually matrices in color space.
At lowest order $S_{LI}={\rm Tr}[c_L^{\dag} c_I]$.
As we shall see, the trace of the product of the matrices $H$ and $S$ 
at lowest order gives the Born cross section.
Full details of the resummation formalism for jet production are given in 
Refs.~\cite{Kidonakis:1998bk,Kidonakis:1998nf,Laenen:1998qw,
Kidonakis:1998qe,Kidonakis:2000ze}.

The moments of the perturbative partonic cross section are then given by
\beqa
E_J \frac{d^3{\hat \sigma}_{f_a f_b \rightarrow J X}(N)} {d^3  p_J} 
& \equiv &  \int \frac{ds_4}{s} \, {\rm e}^{-Ns_4/s} E_J
\frac{d^3{\hat \sigma}_{f_af_b\rightarrow J X}(s_4)}{d^3 p_J}
\nonumber \\ &&
=\frac{{\tilde{\psi}}_{f_a/f_a}(N_a)\, {\tilde{\psi}}_{f_b/f_b}(N_b)}
{{\tilde{\phi}}_{f_a/f_a}(N_a)\,  {\tilde{\phi}}_{f_b/f_b}(N_b)}
{\tilde{J}}(N){\tilde{J^r}}(N)  \, H \, {\tilde{S}}({p_T}/(N\mu_F)) \, .
\label{sigfact}
\eeqa
where $J$ represents the final-state observed jet,  
and $J^r$ the partons recoiling against the jet.
Note that $N_a=N(-u/s)$ and $N_b=N(-t/s)$, with $N$ the moment variable.
The moments of the plus distributions, Eq.~(\ref{s4plus}), in the cross 
section produce powers of $\ln N$ as high as $\ln^{2n} N$. The leading 
and next-to-leading logarithms of $N$ are then resummed to all orders
in perturbation theory.  

The resummed cross section in moment space follows from the 
resummation of the $N$-dependence of each of the functions in 
Eq.~(\ref{sigfact}), and may be written as
\cite{Kidonakis:1998bk,Laenen:1998qw}
\beqa
E_J \frac{d^3{{\hat \sigma}^{\rm resum}}_{f_a f_b \rightarrow J X}(N)}
{d^3  p_J} &=&\frac{1}{s} \exp \left\{-\sum_{i=a,b} 
2\int_{\mu_F}^{2p_i \cdot \zeta}{d\mu'\over\mu'}\; 
C_{(f_i)}\frac{\alpha_s(\mu'{}^2)}{\pi} \ln N_i  \right.
\nonumber \\ && \hspace{-40mm} \left.
{}-\sum_{i=a,b} \int^1_0 dz \frac{z^{N_i-1}-1}{1-z}\;
\left[\int^1_{(1-z)^2} \frac{d\lambda}{\lambda}
A^{(f_i)}\left[\alpha_s\left(\lambda (2 p_i \cdot \zeta)^2\right)\right]
+\frac{1}{2}\nu^{(f_i)}\left[\alpha_s((1-z)^2(2 p_i \cdot \zeta)^2)\right]
\right] \right\}
\nonumber \\ && \hspace{-40mm}
\times\; \exp \left\{\sum_{j=J,J^r}\int^1_0 dz \frac{z^{N-1}-1}{1-z}\;
\left[\int^{(1-z)}_{(1-z)^2} \frac{d\lambda}{\lambda}
A^{(j)}\left[\alpha_s(\lambda p_T^2)\right]\right.\right.
\nonumber \\ && \left.\left. \hspace{15mm}
{}+B'_{(j)}\left[\alpha_s((1-z) p_T^2) \right]
+B''_{(j)}\left[\alpha_s((1-z)^2 p_T^2) \right]\right] \right\}
\nonumber\\ && \hspace{-40mm} \times\; \exp \left[2\int_{\mu_F}^{p_T} 
{d\mu' \over \mu'} \left(\gamma_a(\alpha_s(\mu'^2))
+\gamma_b(\alpha_s(\mu'^2))\right)\right]
\exp\left[4\int_{\mu_R}^{p_T}\frac{d\mu'}{\mu'} \beta(\alpha_s(\mu'^2))\right]
{\rm Tr} \left\{H^{({\rm f})}\left(\alpha_s(\mu_R^2)\right) \right.
\nonumber\\ && \hspace{-40mm} \times\; \left.
\bar{P} \exp \left[\int_{p_T}^{p_T/N} {d\mu' \over \mu'}\; 
\Gamma_S^{({\rm f})}{}^{\dag}\left(\alpha_s(\mu'^2)\right)\right]\;
{\tilde S}^{({\rm f})} \left(\alpha_s\left(p_T^2/N^2\right) \right) \; 
P \exp \left[\int_{p_T}^{p_T/N} {d\mu' \over \mu'}\; \Gamma_S^{({\rm f})}
\left(\alpha_s(\mu'^2)\right)\right] \right\} \, .
\nonumber \\
\label{resjet}
\eeqa

The first exponent resums the $N$-dependence of the incoming partons 
\cite{Sterman:1987aj,Catani:1989ne}.
At next-to-leading order accuracy in $\ln N$,
$A^{(f)}(\alpha_s) = C_f [(\alpha_s/\pi)
+(K/2)(\alpha_s/\pi)^2]$,
and $\nu^{(f)}=2C_f (\alpha_s/\pi)$.
Here $C_f=C_F=(N_c^2-1)/(2N_c)$ for an incoming quark,
and $C_f=C_A=N_c$ for an incoming gluon, with $N_c$ the number of colors,
while $K= C_A (67/18-\pi^2/6)-5 n_f/9$,
where $n_f$ is the number of quark flavors.
Also $\zeta^{\mu}=p_J^{\mu}/p_T$.
Note that the leading contributions from this exponent enhance
the cross section \cite{Kidonakis:1996hb}.

The second exponent resums the $N$-dependence
of the final-state observed jet, $J$, and of partons $J^r$ recoiling against 
the jet \cite{Kidonakis:1998bk,Laenen:1998qw,Kidonakis:2000ur}, with
$B'_{(q)}=(\alpha_s/\pi)(-3C_F/4)$,
$B''_{(q)}=(\alpha_s/\pi) C_F [\ln(2\nu_q)-1]$,
$B'_{(g)}=(\alpha_s/\pi)(-\beta_0/4)$, and
$B''_{(g)}=(\alpha_s/\pi) C_A [\ln(2\nu_g)-1]$ 
\cite{Laenen:1998qw,Kidonakis:2000ur,Kidonakis:2000hq}.
The $\nu_i$ terms are gauge dependent. They are defined by
$\nu_i \equiv (\beta_i \cdot n)^2/|n|^2$,
where $\beta_i=p_i {\sqrt {2/s}}$ are the particle velocities
and $n$ is the axial gauge vector, chosen so that $p_i\cdot \zeta=p_i \cdot n$
for $i=a,b$ \cite{Laenen:1998qw,Kidonakis:2000ur} and $|n|^2=1$.
Also $\beta_0= (11C_A-2n_f)/3$ is the one-loop coefficient of the $\beta$
function, $\beta(\alpha_s) \equiv \mu d \ln g/d \mu
=-\beta_0 \alpha_s/(4 \pi) +...$. 
Note that this exponent contributes a suppresion to the cross section
at leading order, in contrast
to the first exponent. This suppresion depends crucially on the way
the cross section is constructed as discussed in Ref. \cite{Kidonakis:1998bk}.
 
In the third exponent, $\gamma_q=(\alpha_s/\pi)(3C_F/4)$ and 
$\gamma_g=(\alpha_s/\pi)(\beta_0/4)$ are anomalous dimensions of 
the fields $\psi$ for quarks and gluons, respectively. 

The trace in Eq.~(\ref{resjet}) is taken in color space.
The evolution of the soft function from scale $p_T/N$ to $p_T$
is given in terms of the matrix $\Gamma_S$, the soft anomalous 
dimension \cite{Kidonakis:1998nf,Kidonakis:1997gm}. 
The symbols $P$ and ${\bar P}$ denote path ordering in the same sense
as the variable $\mu'$ and against it, respectively.
The superscript $({\rm f})$ on $H$, $S$, and $\Gamma_S$ stands for the
process $f_a f_b \rightarrow JX$.
The soft anomalous dimension matrix, evaluated through the calculation of
one-loop eikonal vertex corrections, has been presented for all partonic
processes in jet production in Ref. \cite{Kidonakis:1998nf}. It can be 
written in the form 
\beq
(\Gamma^{({\rm f})}_S)_{KL}=(\Gamma^{({\rm f})}_{S'})_{KL}
+ \delta_{KL}  \frac{\alpha_s}{\pi} \sum_{i=a,b,1,2}C_{(f_i)} \, 
\frac{1}{2} \,  [-\ln(2\nu_i)+1-\pi i] \, ,
\label{gammagaug}
\eeq
where the gauge-dependent terms appear explicitly. 
We note that all gauge dependence cancels out in the cross section.

In a color basis where the soft anomalous dimension $\Gamma_S$
is diagonal, the path-ordered exponentials of matrices in 
Eq. (\ref{resjet}) reduce
to simple exponentials; however, in practice the diagonalization
procedure is complicated \cite{Kidonakis:2000ze}.

Note that the Born cross section is given simply by the trace of
the lowest-order soft and hard functions as
\beq  
E_J \frac{d^3{{\hat \sigma}^B}_{f_a f_b \rightarrow J X}}
{d^3 p_J} =\frac{1}{s} {\rm Tr}[H^{f_a f_b \rightarrow J X}
S^{f_a f_b \rightarrow J X}] \, . 
\label{born}
\eeq

Since the resummed cross section is given in moment space, one
needs to perform an inverse transform to calculate the physical
cross section. A prescription is then needed to take care of the
regions of phase space where the running coupling diverges, and the 
results do depend on the prescription. 
The resummed cross section may also be expanded in powers of $\alpha_s$ 
and then one can perform the inversion order-by-order, thus avoiding
the use of a prescription as well as the diagonalization procedure
mentioned above. We prefer the latter approach, and 
expand the resummed cross section through next-to-next-to-leading order 
at NLL accuracy. Thus, we present the first calculation
of NNLO-NLL corrections for jet production.

In the appendices we present the matrices for the
lowest-order hard and soft functions 
(some of which have appeared in Ref.~\cite{Oderda:2000kr} 
in the context of dijet rapidity gaps) and
the one-loop soft anomalous dimensions $\Gamma_{S'}$ 
\cite{Kidonakis:1998nf,Kidonakis:1998qe, Kidonakis:2000ze}
for each partonic process in a given color basis.
The Born cross section, calculated using Eq.~(\ref{born}),
is also given; it agrees with well-known results 
\cite{Cutler:1978qm,Ellis:1986er,Aversa:1989vb},
thus providing a check on the lowest-order hard and soft matrices.
We then calculate the NLO and NNLO threshold corrections to the single-jet
inclusive cross section from the one- and two-loop expansions of the 
resummed cross section.
In the following section we apply those results to high-$E_T$ jet
production at the Tevatron.

\mysection{Jet production at the Tevatron}

In this section the NNLO predictions are compared to data for the single jet 
inclusive cross section at the Tevatron. The NLO predictions have been
generated using the EKS jet program
\cite{Ellis:1989vm,Ellis:1990ek,Kunszt:1992tn} with the CTEQ5M parton
distributions \cite{Lai:2000wy}. The NNLO corrections have been calculated in
accordance with the expressions contained in Appendices A-E and added to the
NLO predictions. In all the cases shown here, the renormalization and
factorization scales have been set equal to each other.

\begin{figure}
\centerline{
\psfig{file=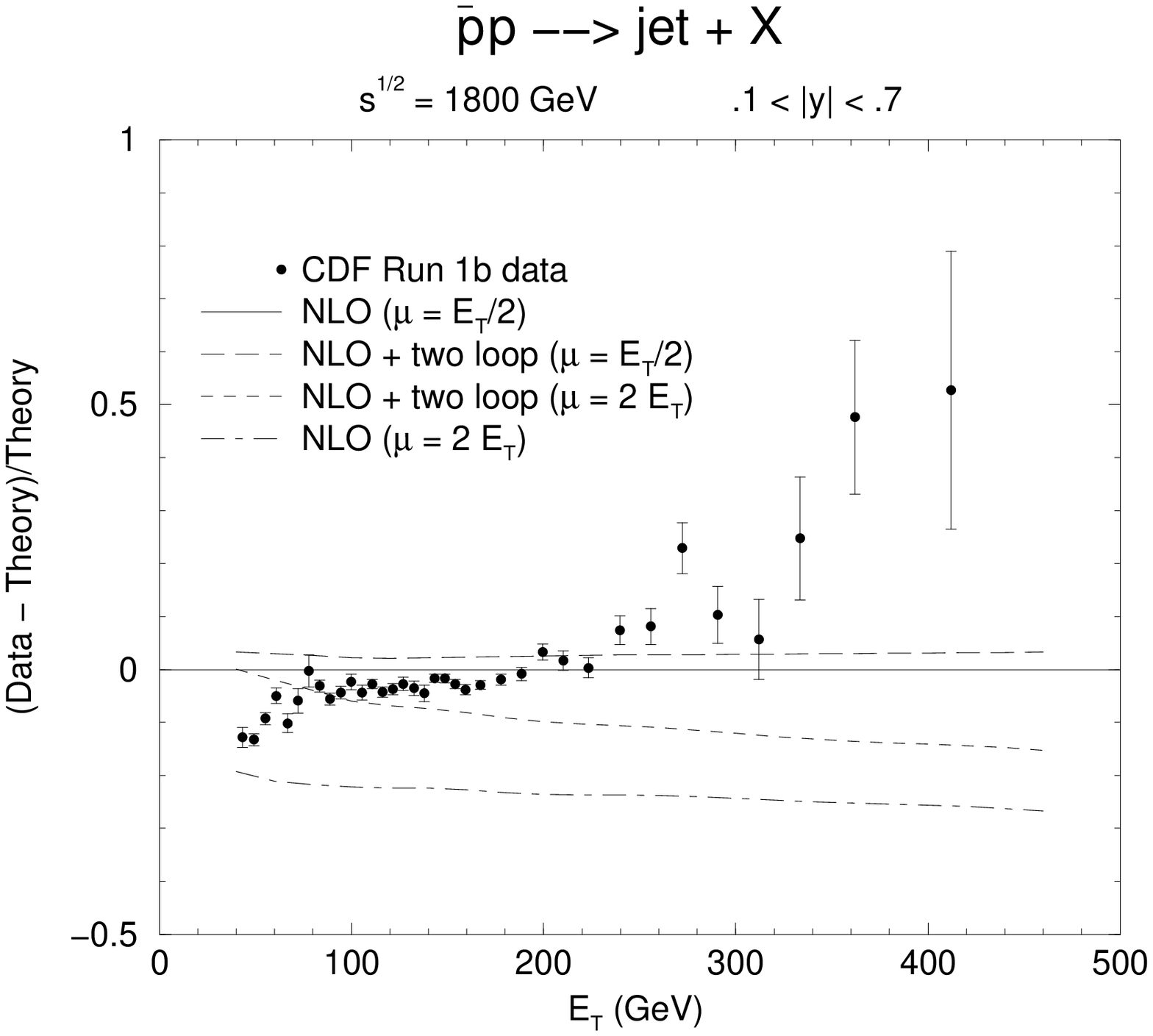,height=3in,width=4in,clip=}}
{Figure 1. NLO and NLO + NNLO results for inclusive jet production
at the Tevatron compared to data from the CDF Collaboration
\cite{CDF}.}
\label{fig1}
\end{figure}
 
In Fig. 1 CDF data \cite{CDF} are shown compared to the NLO predictions with
a scale of $E_T/2$ in the form of (data-theory)/theory. The excess at high
$E_T$ relative to the lower $E_T$ results is clear. The dash-dot curve shows
the effect of using 2 $E_T$ as the scale. The scale dependence over the $E_T$
range shown varies between about 20\% and 28\%. Next, the effect of 
adding in the two loop corrections with the scale set to $E_T/2$ is shown by 
the long dashed line, while the corresponding curve for $2 E_T$ is shown by the
short dashed line. The relative spacing between the long and short dashed 
lines 
as compared to the solid and dash-dot lines shows that the inclusion of the two
loop terms has reduced the scale dependence. In addition, the two loop
corrections are rather small for the scale choice of $E_T/2$. These results are
very similar to those obtained in Ref. \cite{Kidonakis:2000hq} for direct
photon production. Using the same notation as in Fig. 1, a comparison with 
D\O\  jet data \cite{Abbott:1999ya} is shown in Fig. 2. 
As is the case with the CDF data in
Fig. 1, the D\O\  data show a definite slope with respect to the theory. In 
fact, 
allowing for a small shift in overall normalization, the deviations of the
theory from the data are quite consistent for the two data sets.
In both cases it is clear that the 
threshold corrections are not sufficient to increase the slope of the
theoretical predictions to match that of the data - the high $E_T$ excess is
still apparent.

\begin{figure}
\centerline{
\psfig{file=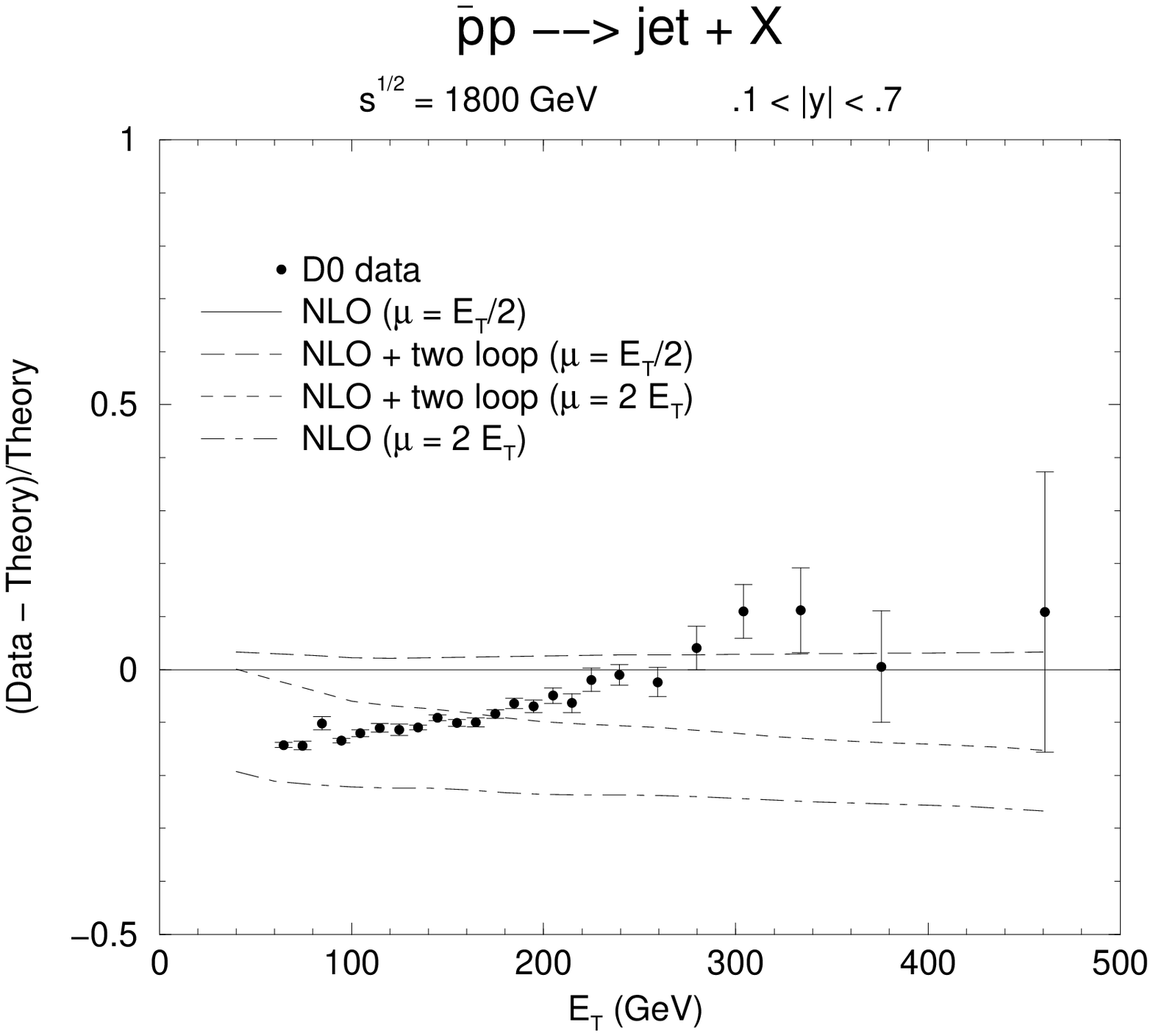,height=3in,width=4in,clip=}}
{Figure 2. NLO and NLO + NNLO results for inclusive jet production
at the Tevatron compared to data from the D\O\ Collaboration
\cite{Abbott:1999ya}.}
\label{fig2}
\end{figure}

\begin{figure}
\centerline{
\psfig{file=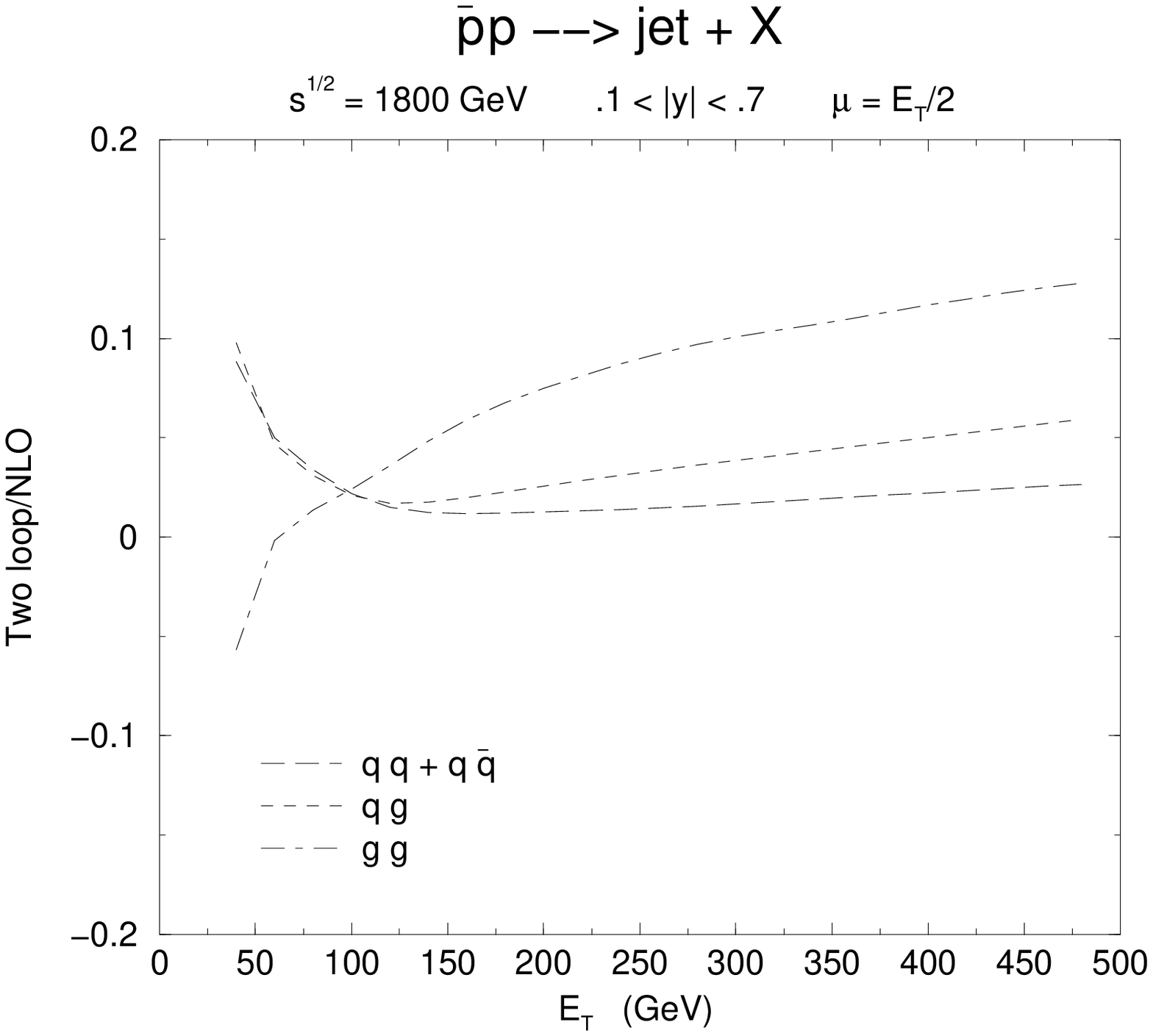,height=3in,width=4in,clip=}}
{Figure 3. Ratio of the NNLO corrections to the NLO calculations for various 
subprocesses.}
\label{fig3}
\end{figure}

It is normally expected that the effects of the threshold resummation
corrections should increase with increasing $E_T$ since the steeply falling 
parton distributions restrict the values of $s_4$ to be near the 
parton-parton scattering threshold where the corrections are large. However, 
the 
results shown in Figs. 1 and 2 show relatively flat corrections. To understand
why this is so, the ratio of the NNLO correction to the NLO calculation is
shown in Fig. 3 for the three types of subprocesses - quark-quark (or 
antiquark), quark-gluon, and gluon-gluon. In each case the ratio shows a rise 
throughout the high $E_T$ region. At high values of $E_T$ the corrections to
the gluon-gluon subprocess are largest, followed by those for the 
quark-gluon and 
quark-quark subprocesses. This ordering is in accord with the size of the color
factors associated with the dominant $(\ln^3(s_4/p_T^2))/s_4$ term. 
Next, in Fig. 4 the relative contributions
to the total jet rate from the same three classes of subprocesses are shown.
As $E_T$ increases, the gluon-gluon subprocess falls rapidly, followed by the
quark-gluon contributions, so that the quark-quark terms dominate at the 
highest $E_T$. Comparing Figs. 3 and 4, one can see that as $E_T$ increases, 
the large gluon-gluon
corrections contribute a decreasing share to the total NNLO correction 
while the share from the smaller quark-quark corrections increases, thereby 
yielding the relatively flat behavior shown in Figs. 1 and 2. Predictions for 
a center-of-mass energy of 2 TeV are shown in Fig. 5 where the same behavior 
is seen to extend past $E_T$=600 GeV. We have examined similar predictions for
the LHC at 14 TeV and the same pattern is seen there, as well. 

\begin{figure}
\centerline{
\psfig{file=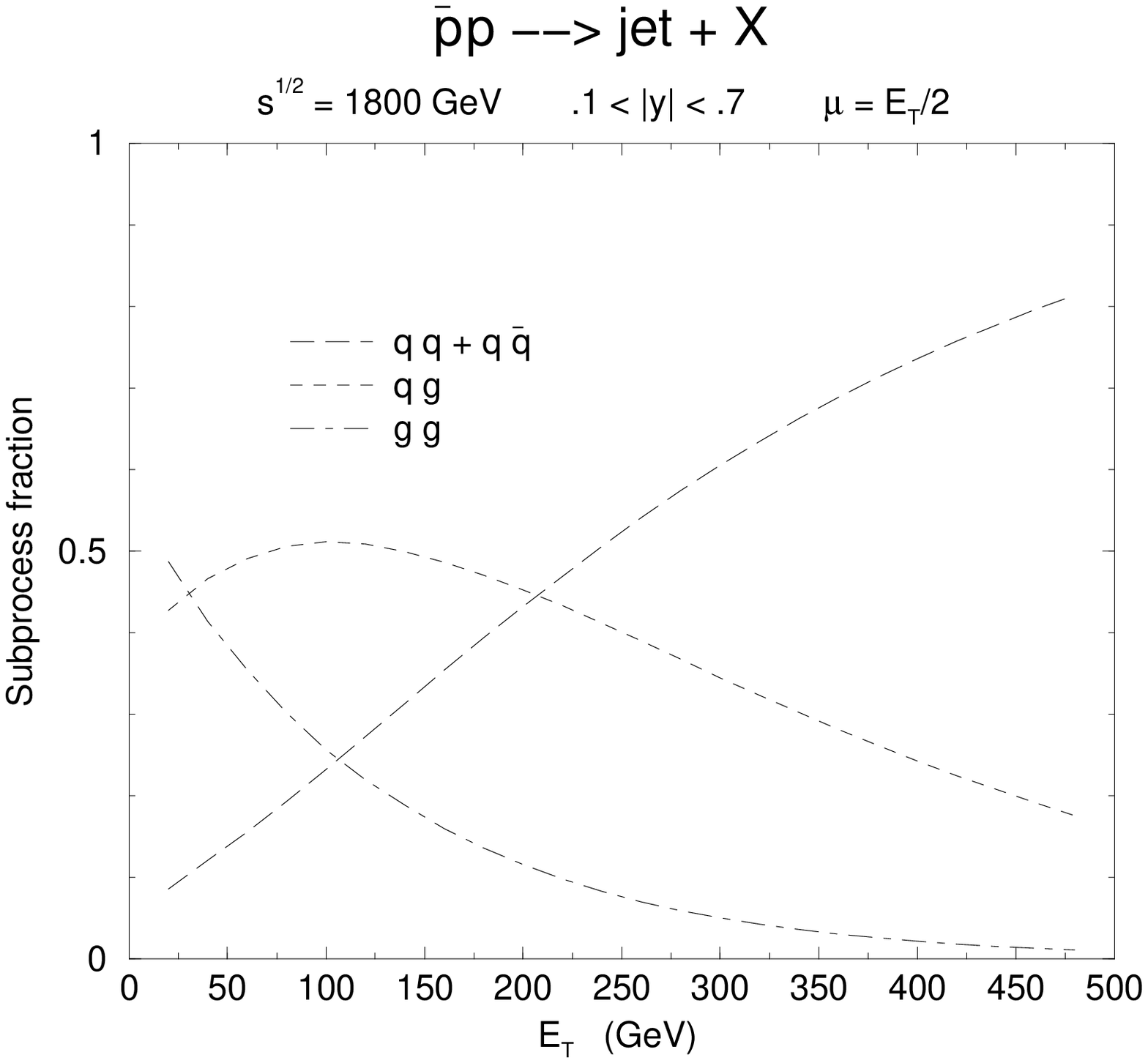,height=3in,width=4in,clip=}}
{Figure 4. Relative contributions of various subprocesses to the NLO 
calculation of inclusive jet production.}
\label{fig4}
\end{figure}

\begin{figure}
\centerline{
\psfig{file=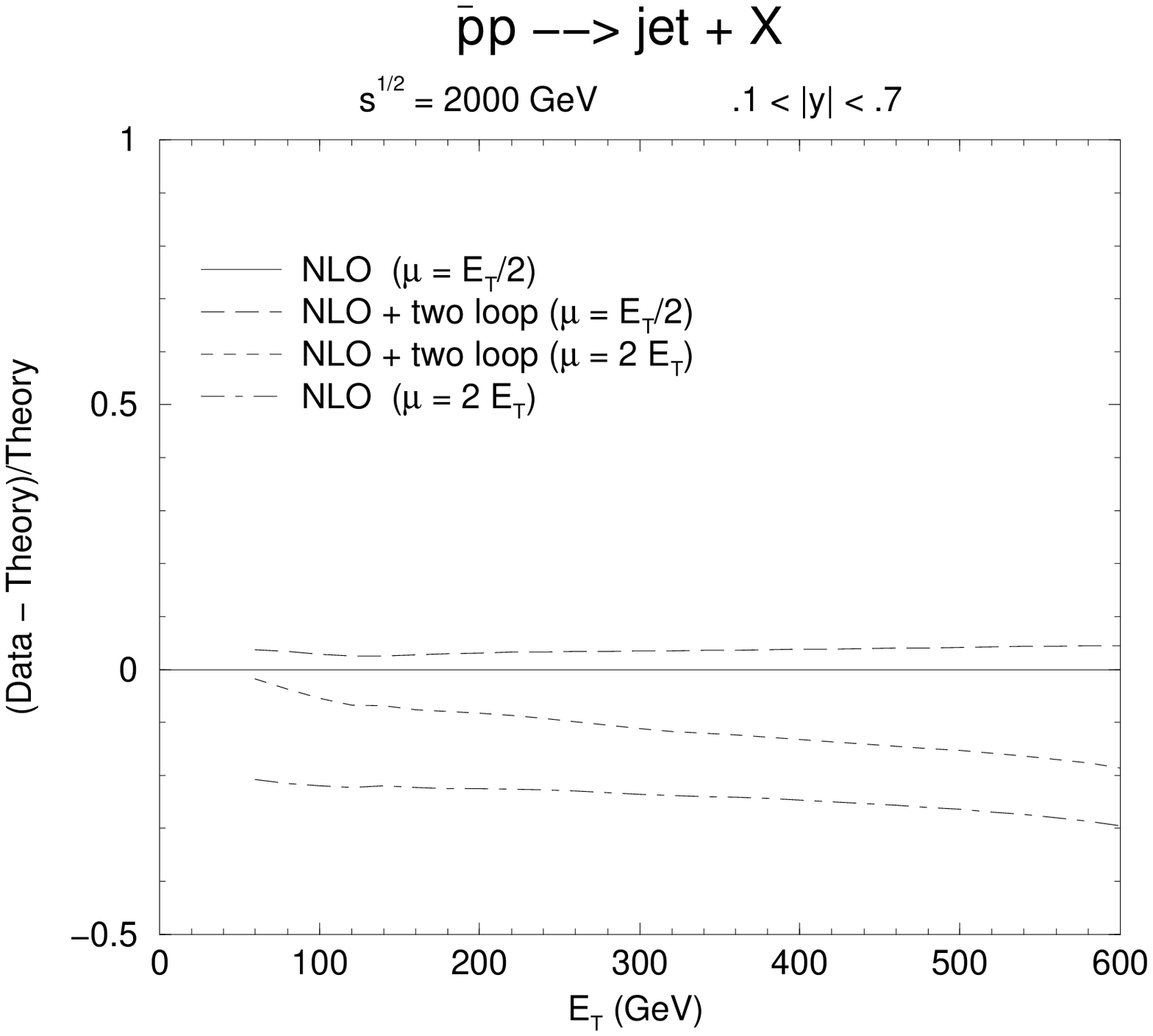,height=3in,width=4in,clip=}}
{Figure 5. Predictions for the NNLO corrections to inclusive jet production 
at a center-of-mass energy of 2 TeV.}
\label{fig5}
\end{figure}

   In Ref. \cite{Catani:1996yz} threshold resummation corrections 
at leading logarithmic accuracy to dijet 
production in hadronic collisions were considered. There it was found that the 
the corrections to the dijet mass distribution were less than about 5\% in the 
kinematic range considered here when a scale $\mu = M_{dijet}/2$ was used. 
This is consistent with our findings, although the details of the corrections 
and some of the techniques used differ in the two cases. 

\mysection{Conclusion}

In this paper the corrections to existing NLO calculations from threshold 
resummation in jet production in hadron-hadron collisions have been studied 
using the formalism of Refs. \cite{Kidonakis:1998bk, Kidonakis:1998nf, 
Laenen:1998qw}. Detailed expressions for the NNLO
corections to NLL accuracy have been presented for each of the parton-parton
scattering processes. It is found that adding the NNLO corrections to NLO 
predictions results in a decrease in the scale dependence and that the 
corrections are rather small if the renormalization and factorization scales
are chosen to $E_T/2$. The NNLO corrections are smallest for quark-quark
subprocesses and increase in size for the quark-gluon and gluon-gluon
subprocesses. However, the more rapid decrease with increasing $E_T$ for the
latter subprocesses results in a relatively flat behavior for the NNLO 
corrections as $E_T$ increases. It therefore appears that threshold 
resummation 
is not sufficient to remove the remaining discrepancies between the data and
the predictions of QCD for high $E_T$ jet production.

\mysection*{Acknowledgements}

The authors wish to thank Gianluca Oderda and George Sterman 
for very useful discussions and correspondence. 
This work was supported in part by the U.S. 
Department of Energy.

\appendix

\mysection{NNLO-NLL corrections for quark-antiquark annihilation processes}

We begin with the quark-antiquark annihilation processes,
\beq
q\left(p_a, r_a \right)+\bar{q}\left(p_b, r_b \right) \rightarrow
q\left(p_1, r_1 \right)+\bar{q}\left(p_2, r_2 \right) \, ,
\eeq
where the $p_i$ and $r_i$ are momentum and color labels, respectively.  
We use the notation
\beq
T\equiv \ln\left(\frac{-t}{s}\right)+\pi i \, , \quad \quad
U\equiv \ln\left(\frac{-u}{s}\right)+\pi i \, ,
\label{eq:new2form}
\eeq
where
\beq
s=\left(p_a+p_b \right)^2 \, , \quad t=\left(p_a-p_1 \right)^2 \, , 
\quad u=\left(p_b-p_1 \right)^2 \, ,
\label{Mandlst}
\eeq
are the usual Mandelstam invariants.
In the $t$-channel singlet-octet color basis
\beq
c_1^{q {\bar q}\rightarrow q {\bar q}}=\delta_{r_a r_1}\delta_{r_b r_2} \, ,
\quad
c_2^{q {\bar q}\rightarrow q {\bar q}}=(T_F^c)_{r_1 r_a}(T_F^c)_{r_b r_2}\, ,
\eeq
where the $T_F^c$ are the generators of $SU(3)$ in the fundamental
representation, the soft matrix $S$ at lowest order, with elements
$S_{LI}={\rm Tr}[c_L^{\dag} c_I]$, is given by 
\beq
S^{q {\bar q}\rightarrow q {\bar q}}=\left[
                \begin{array}{cc}
                 N_c^2 & 0  
\vspace{2mm} \\
                 0 & (N_c^2-1)/4
               \end{array} \right] \, ,
\eeq
and the one-loop soft anomalous dimension matrix $\Gamma_{S'}$ is 
\cite{Kidonakis:1998nf,Kidonakis:1997gm,Botts:1989kf}
\beq
\Gamma_{S'}^{q {\bar q}\rightarrow q {\bar q}}=\frac{\alpha_s}{\pi}\left[
                \begin{array}{cc}
                 2{C_F} T  &   -\frac{C_F}{N_c} U  \vspace{2mm} \\
                -2U    &-\frac{1}{N_c}(T-2 U)
                \end{array} \right]\, .
\eeq
 
There are three different quark-antiquark subprocesses to consider
depending on the quark flavors. 

\subsection{NNLO corrections for $q_j {\bar q}_j \rightarrow q_j {\bar q}_j $}

The hard matrix for the process $q_j {\bar q}_j \rightarrow q_j {\bar q}_j$
at lowest order, whose elements come from the squares of the 
color-decomposed tree amplitudes, is given by 
(see also Ref.~\cite{Oderda:2000kr})
\beq
H^{q_j {\bar q_j}\rightarrow q_j {\bar q_j}}=\alpha_s^2 \left[
                \begin{array}{cc}
                 H_{11}^{q_j {\bar q_j}\rightarrow q_j {\bar q_j}}
  & H_{12}^{q_j {\bar q_j}\rightarrow q_j {\bar q_j}}  
\vspace{2mm} \\
                 H_{12}^{q_j {\bar q_j}\rightarrow q_j {\bar q_j}}
  & H_{22}^{q_j {\bar q_j}\rightarrow q_j {\bar q_j}}
               \end{array} \right] \, ,
\eeq
with
\beqa
H_{11}^{q_j {\bar q_j}\rightarrow q_j {\bar q_j}}
&=&\frac{2C_F^2}{N_c^4} \frac{(t^2+u^2)}{s^2} \, ,
\nonumber \\
H_{12}^{q_j {\bar q_j}\rightarrow q_j {\bar q_j}}
&=&\frac{2C_F}{N_c^3} \left[-\frac{(t^2+u^2)}{N_c s^2}
 +\frac{u^2}{st}\right] \, ,
\nonumber \\
H_{22}^{q_j {\bar q_j}\rightarrow q_j {\bar q_j}}
&=&\frac{1}{N_c^2} \left[\frac{2}{N_c^2}\frac{(t^2+u^2)}{s^2}
+2\frac{(s^2+u^2)}{t^2}-\frac{4}{N_c}\frac{u^2}{st}\right] \, .
\eeqa

The Born cross section is given, using Eq. (\ref{born}), by
\beq
E_J \frac{d^3{\hat \sigma}^B_{q_j {\bar q}_j \rightarrow q_j {\bar q}_j}}
{d^3p_J}\equiv \sigma^B_{q_j {\bar q}_j \rightarrow q_j {\bar q}_j} \delta(s_4)
=\alpha_s^2 \frac{(N_c^2-1)}{2N_c^2 \, s} \left[\frac{t^2+u^2}{s^2}
+\frac{s^2+u^2}{t^2}-\frac{2}{N_c}\frac{u^2}{st}\right] \delta(s_4) \, .
\eeq

The NLO threshold corrections from the one-loop expansion of the 
resummed cross section at NLL accuracy are then
\beqa
E_J\frac{d^3{\hat \sigma}^{(1)}_{q_j {\bar q}_j \rightarrow q_j {\bar q}_j}}
{d^3p_J}&=&\frac{\alpha_s}{\pi}
\sigma^B_{q_j {\bar q}_j \rightarrow q_j {\bar q}_j}
\left\{2C_F \left[\frac{\ln(s_4/p_T^2)}{s_4}\right]_+ \right.
\nonumber \\ &&  \hspace{-20mm} \left.
{}+\left[-2C_F\ln\left(\frac{\mu_F^2}{p_T^2}\right)
-\frac{3}{2}C_F-\frac{(N_c^2+1)}{N_c}\ln\left(\frac{-t}{s}\right)
-\frac{(N_c^2-5)}{N_c}\ln\left(\frac{-u}{s}\right)\right]
\left[\frac{1}{s_4}\right]_+ \right\}
\nonumber \\ && \hspace{-25mm}
{}+\frac{\alpha_s^3}{\pi}\left\{\frac{4C_F^2}{N_c}\ln\left(\frac{-t}{s}\right)
\frac{(t^2+u^2)}{s^2}-\frac{8C_F^2}{N_c^2}\ln\left(\frac{-u}{s}\right)
\frac{u^2}{st}\right\} \left[\frac{1}{s_4}\right]_+
\nonumber \\ && \hspace{-25mm}
{}+\frac{\alpha_s}{\pi}
\sigma^B_{q_j {\bar q}_j \rightarrow q_j {\bar q}_j} \delta(s_4)
\left\{-C_F\left[\ln\left(\frac{p_T^2}{s}\right)+\frac{3}{2}\right]
\ln\left(\frac{\mu_F^2}{p_T^2}\right)
+\frac{\beta_0}{2}\ln\left(\frac{\mu_R^2}{p_T^2}\right)\right\} \, .
\eeqa

From the two-loop expansion of the resummed cross section at NLL accuracy, 
we obtain the following NNLO threshold corrections 
\beqa
E_J\frac{d^3{\hat \sigma}^{(2)}_{q_j {\bar q}_j \rightarrow q_j {\bar q}_j}}
{d^3p_J}&=&\left(\frac{\alpha_s}{\pi}\right)^2
\sigma^B_{q_j {\bar q}_j \rightarrow q_j {\bar q}_j}
\left\{2C_F^2 \left[\frac{\ln^3(s_4/p_T^2)}{s_4}\right]_+ \right.
\nonumber \\ &&  \hspace{-10mm} 
{}+3C_F\left[-2C_F\ln\left(\frac{\mu_F^2}{p_T^2}\right)
-\frac{3}{2}C_F-\frac{(N_c^2+1)}{N_c}\ln\left(\frac{-t}{s}\right) \right.
\nonumber \\ && \quad \quad \left. \left.
-\frac{(N_c^2-5)}{N_c}\ln\left(\frac{-u}{s}\right)-\frac{\beta_0}{12}\right]
\left[\frac{\ln^2(s_4/p_T^2)}{s_4}\right]_+ \right\}
\nonumber \\ &&  \hspace{-20mm} 
{}+\frac{\alpha_s^4}{\pi^2} \frac{12C_F^3}{N_c}
\left[\ln\left(\frac{-t}{s}\right)
\frac{(t^2+u^2)}{s^2}-\frac{2}{N_c}\ln\left(\frac{-u}{s}\right)
\frac{u^2}{st}\right] \left[\frac{\ln^2(s_4/p_T^2)}{s_4}\right]_+
\nonumber \\ &&  \hspace{-20mm} 
{}+\left(\frac{\alpha_s}{\pi}\right)^2
\sigma^B_{q_j {\bar q}_j \rightarrow q_j {\bar q}_j} 
\left\{4C_F\left[-\frac{1}{2}C_F\ln\left(\frac{p_T^2}{s}\right)
+\frac{3}{4}C_F+\frac{(N_c^2+1)}{N_c}\ln\left(\frac{-t}{s}\right)\right.\right.
\nonumber \\ && \hspace{-15mm} \left. \left.
{}+\frac{(N_c^2-5)}{N_c}\ln\left(\frac{-u}{s}\right)
+C_F\ln\left(\frac{\mu_F^2}{p_T^2}\right)\right]
\ln\left(\frac{\mu_F^2}{p_T^2}\right)
+\frac{3}{2}C_F\beta_0\ln\left(\frac{\mu_R^2}{p_T^2}\right)\right\}
\left[\frac{\ln(s_4/p_T^2)}{s_4}\right]_+ 
\nonumber \\ &&  \hspace{-20mm}
{}+\frac{\alpha_s^4}{\pi^2}\frac{16C_F^3}{N_c}
\left[-\ln\left(\frac{-t}{s}\right)
\frac{(t^2+u^2)}{s^2}+\frac{2}{N_c}\ln\left(\frac{-u}{s}\right)
\frac{u^2}{st}\right]\ln\left(\frac{\mu_F^2}{p_T^2}\right) 
\left[\frac{\ln(s_4/p_T^2)}{s_4}\right]_+
\nonumber \\ &&  \hspace{-20mm} 
{}+\left(\frac{\alpha_s}{\pi}\right)^2
\sigma^B_{q_j {\bar q}_j \rightarrow q_j {\bar q}_j}  \left\{ C_F
\left[2C_F\ln\left(\frac{p_T^2}{s}\right)+3C_F+\frac{\beta_0}{4}\right]
\ln^2\left(\frac{\mu_F^2}{p_T^2}\right) \right.
\nonumber \\ &&  \hspace{15mm} \left.
{}-\frac{3}{2} \beta_0 C_F \ln\left(\frac{\mu_F^2}{p_T^2}\right) 
\ln\left(\frac{\mu_R^2}{p_T^2}\right)\right\}
\left[\frac{1}{s_4}\right]_+ \, .
\eeqa

\subsection{NNLO corrections for $q_j {\bar q}_j \rightarrow q_k {\bar q}_k $}

The hard matrix at lowest order is 
\beq
H^{q_j {\bar q_j}\rightarrow q_k {\bar q_k}}=\alpha_s^2 \left[
                \begin{array}{cc}
                 (C_F^2/N_c^2)h^{q_j {\bar q_j}\rightarrow q_k {\bar q_k}}
 & -(C_F/N_c^2)h^{q_j {\bar q_j}\rightarrow q_k {\bar q_k}}  
\vspace{2mm} \\
                 -(C_F/N_c^2)h^{q_j {\bar q_j}\rightarrow q_k {\bar q_k}}
  & h^{q_j {\bar q_j}\rightarrow q_k {\bar q_k}}/N_c^2
               \end{array} \right] \, ,
\eeq
with
\beq
h^{q_j {\bar q_j}\rightarrow q_k {\bar q_k}}
=\frac{2}{N_c^2} \frac{(t^2+u^2)}{s^2} \, .
\eeq

The Born cross section is
\beq
E_J \frac{d^3{\hat \sigma}^B_{q_j {\bar q}_j \rightarrow q_k {\bar q}_k}}
{d^3p_J}\equiv \sigma^B_{q_j {\bar q}_j \rightarrow q_k {\bar q}_k} \delta(s_4)
=\alpha_s^2 \frac{(N_c^2-1)}{2N_c^2\, s} \frac{(t^2+u^2)}{s^2} \delta(s_4) \, .
\eeq

The NLO corrections are
\beqa
E_J\frac{d^3{\hat \sigma}^{(1)}_{q_j {\bar q}_j \rightarrow q_k {\bar q}_k}}
{d^3p_J}&=&\frac{\alpha_s}{\pi}
\sigma^B_{q_j {\bar q}_j \rightarrow q_k {\bar q}_k}
\left\{2C_F \left[\frac{\ln(s_4/p_T^2)}{s_4}\right]_+ \right.
\nonumber \\ &&  \hspace{-25mm} \left.
{}+\left[-2C_F\ln\left(\frac{\mu_F^2}{p_T^2}\right)
-\frac{3}{2}C_F+\frac{(N_c^2-3)}{N_c}\ln\left(\frac{-t}{s}\right)
-\frac{(N_c^2-5)}{N_c}\ln\left(\frac{-u}{s}\right)\right]
\left[\frac{1}{s_4}\right]_+ \right.
\nonumber \\ && \hspace{-25mm} \left.
{}+\delta(s_4)
\left[-C_F\left(\ln\left(\frac{p_T^2}{s}\right)+\frac{3}{2}\right)
\ln\left(\frac{\mu_F^2}{p_T^2}\right)
+\frac{\beta_0}{2}\ln\left(\frac{\mu_R^2}{p_T^2}\right)\right]\right\} \, .
\eeqa

The NNLO corrections are
\beqa
E_J\frac{d^3{\hat \sigma}^{(2)}_{q_j {\bar q}_j \rightarrow q_k {\bar q}_k}}
{d^3p_J}&=&\left(\frac{\alpha_s}{\pi}\right)^2
\sigma^B_{q_j {\bar q}_j \rightarrow q_k {\bar q}_k}
\left\{2C_F^2 \left[\frac{\ln^3(s_4/p_T^2)}{s_4}\right]_+ \right.
\nonumber \\ &&  \hspace{-20mm}
{}+3C_F\left[-2C_F\ln\left(\frac{\mu_F^2}{p_T^2}\right)
-\frac{3}{2}C_F+\frac{(N_c^2-3)}{N_c}\ln\left(\frac{-t}{s}\right) \right.
\nonumber \\ &&  \hspace{-5mm} \left.
{}-\frac{(N_c^2-5)}{N_c}\ln\left(\frac{-u}{s}\right)-\frac{\beta_0}{12}\right]
\left[\frac{\ln^2(s_4/p_T^2)}{s_4}\right]_+
\nonumber \\ &&  \hspace{-20mm} 
{}+4C_F\left[-\frac{1}{2}C_F\ln\left(\frac{p_T^2}{s}\right)
+\frac{3}{4}C_F-\frac{(N_c^2-3)}{N_c}\ln\left(\frac{-t}{s}\right)\right.
\nonumber \\ &&  \hspace{-5mm} \left. 
{}+\frac{(N_c^2-5)}{N_c}\ln\left(\frac{-u}{s}\right)
+C_F\ln\left(\frac{\mu_F^2}{p_T^2}\right)\right]
\ln\left(\frac{\mu_F^2}{p_T^2}\right)
\left[\frac{\ln(s_4/p_T^2)}{s_4}\right]_+
\nonumber \\ &&  \hspace{-20mm}
{}+\frac{3}{2}C_F\beta_0\ln\left(\frac{\mu_R^2}{p_T^2}\right)
\left[\frac{\ln(s_4/p_T^2)}{s_4}\right]_+
\nonumber \\ &&  \hspace{-20mm}
{}+C_F\left[2C_F\ln\left(\frac{p_T^2}{s}\right)+3C_F+\frac{\beta_0}{4}\right]
\ln^2\left(\frac{\mu_F^2}{p_T^2}\right) \left[\frac{1}{s_4}\right]_+
\nonumber \\ &&  \hspace{-20mm} \left.
{}-\frac{3}{2}C_F\beta_0\ln\left(\frac{\mu_F^2}{p_T^2}\right) 
\ln\left(\frac{\mu_R^2}{p_T^2}\right)\left[\frac{1}{s_4}\right]_+ \right\} \, .
\eeqa

\subsection{NNLO corrections for $q_j {\bar q}_k \rightarrow q_j {\bar q}_k $}

The hard matrix at lowest order is
\beq
H^{q_j {\bar q_k}\rightarrow q_j {\bar q_k}}=\alpha_s^2 \left[
                \begin{array}{cc}
                 0 & 0  
\vspace{2mm} \\
                 0 & 2(s^2+u^2)/(N_c^2 t^2)
               \end{array} \right] \, .
\eeq

The Born cross section is
\beq
E_J \frac{d^3{\hat \sigma}^B_{q_j {\bar q}_k \rightarrow q_j {\bar q}_k}}
{d^3p_J}\equiv \sigma^B_{q_j {\bar q}_k \rightarrow q_j {\bar q}_k} \delta(s_4)
=\alpha_s^2 \frac{(N_c^2-1)}{2N_c^2\, s} \frac{(s^2+u^2)}{t^2} \delta(s_4) \, .
\eeq

The NLO corrections are
\beqa
E_J\frac{d^3{\hat \sigma}^{(1)}_{q_j {\bar q}_k \rightarrow q_j {\bar q}_k}}
{d^3p_J}&=&\frac{\alpha_s}{\pi}
\sigma^B_{q_j {\bar q}_k \rightarrow q_j {\bar q}_k}
\left\{2C_F \left[\frac{\ln(s_4/p_T^2)}{s_4}\right]_+ \right.
\nonumber \\ &&  \hspace{-20mm} \left.
{}+\left[-2C_F\ln\left(\frac{\mu_F^2}{p_T^2}\right)
-\frac{3}{2}C_F-\frac{(N_c^2+1)}{N_c}\ln\left(\frac{-t}{s}\right)
-\frac{(N_c^2-5)}{N_c}\ln\left(\frac{-u}{s}\right)\right]
\left[\frac{1}{s_4}\right]_+ \right.
\nonumber \\ && \hspace{-20mm} \left.
{}+\delta(s_4)
\left[-C_F\left(\ln\left(\frac{p_T^2}{s}\right)+\frac{3}{2}\right)
\ln\left(\frac{\mu_F^2}{p_T^2}\right)
+\frac{\beta_0}{2}\ln\left(\frac{\mu_R^2}{p_T^2}\right)\right]\right\}\, .
\eeqa

The NNLO corrections are 
\beqa
E_J\frac{d^3{\hat \sigma}^{(2)}_{q_j {\bar q}_k \rightarrow q_j {\bar q}_k}}
{d^3p_J}&=&\left(\frac{\alpha_s}{\pi}\right)^2
\sigma^B_{q_j {\bar q}_k \rightarrow q_j {\bar q}_k}
\left\{2C_F^2 \left[\frac{\ln^3(s_4/p_T^2)}{s_4}\right]_+ \right.
\nonumber \\ &&  \hspace{-20mm} 
{}+3C_F\left[-2C_F\ln\left(\frac{\mu_F^2}{p_T^2}\right)
-\frac{3}{2}C_F-\frac{(N_c^2+1)}{N_c}\ln\left(\frac{-t}{s}\right) \right.
\nonumber \\ && \hspace{-5mm} \left.
-\frac{(N_c^2-5)}{N_c}\ln\left(\frac{-u}{s}\right)-\frac{\beta_0}{12}\right]
\left[\frac{\ln^2(s_4/p_T^2)}{s_4}\right]_+
\nonumber \\ &&  \hspace{-20mm} 
{}+4C_F\left[-\frac{1}{2}C_F\ln\left(\frac{p_T^2}{s}\right)
+\frac{3}{4}C_F+\frac{(N_c^2+1)}{N_c}\ln\left(\frac{-t}{s}\right) \right.
\nonumber \\ &&  \hspace{-5mm} \left.
{}+\frac{(N_c^2-5)}{N_c}\ln\left(\frac{-u}{s}\right)
+C_F\ln\left(\frac{\mu_F^2}{p_T^2}\right)\right]
\ln\left(\frac{\mu_F^2}{p_T^2}\right)
\left[\frac{\ln(s_4/p_T^2)}{s_4}\right]_+
\nonumber \\ &&  \hspace{-20mm}
{}+\frac{3}{2}C_F\beta_0\ln\left(\frac{\mu_R^2}{p_T^2}\right)
\left[\frac{\ln(s_4/p_T^2)}{s_4}\right]_+
\nonumber \\ &&  \hspace{-20mm}
{}+C_F\left[2C_F\ln\left(\frac{p_T^2}{s}\right)+3C_F+\frac{\beta_0}{4}\right]
\ln^2\left(\frac{\mu_F^2}{p_T^2}\right) \left[\frac{1}{s_4}\right]_+
\nonumber \\ &&  \hspace{-20mm} \left.
{}-\frac{3}{2}C_F\beta_0\ln\left(\frac{\mu_F^2}{p_T^2}\right) 
\ln\left(\frac{\mu_R^2}{p_T^2}\right)\left[\frac{1}{s_4}\right]_+ \right\}\, .
\eeqa

\mysection{NNLO-NLL corrections for quark-quark scattering processes}

Next, we analyze quark-quark scattering processes,
\beq 
q\left(p_a, r_a \right)+q\left(p_b, r_b \right) \rightarrow
q\left(p_1, r_1 \right)+q\left(p_2, r_2 \right) \, .
\eeq
In the $t$-channel singlet-octet color basis
\beq
c_1^{qq \rightarrow qq}=(T_F^c)_{r_1 r_a}(T_F^c)_{r_2 r_b}  \, ,
\quad
c_2^{qq \rightarrow qq}=\delta_{r_a r_1} \delta_{r_b r_2} \, ,
\eeq
the soft matrix is given at lowest order by 
\beq
S^{qq \rightarrow qq}=\left[
                \begin{array}{cc}
                 (N_c^2-1)/4 & 0  
\vspace{2mm} \\
                 0 & N_c^2
               \end{array} \right] \, ,
\eeq
and the one-loop soft anomalous dimension matrix is
\cite{Kidonakis:1998nf,Botts:1989kf}
\beq
\Gamma_{S'}^{qq \rightarrow qq}=\frac{\alpha_s}{\pi}\left[
                \begin{array}{cc}
                -\frac{1}{N_c}(T+U)+2C_F U  &  2 U \vspace{2mm} \\
                 \frac{C_F}{N_c} U    & 2{C_F} T
                \end{array} \right].
\eeq

There are two different quark-quark processes to consider
depending on the quark flavors. 

\subsection{NNLO corrections for $q_j q_j \rightarrow q_j q_j $}

The hard matrix for this process is given at lowest order by
\beq
H^{q_j q_j \rightarrow q_j q_j}=\alpha_s^2 \left[
                \begin{array}{cc}
                 H_{11}^{q_j q_j \rightarrow q_j q_j}
 & H_{12}^{q_j q_j \rightarrow q_j q_j}  
\vspace{2mm} \\
                 H_{12}^{q_j q_j \rightarrow q_j q_j}
 & H_{22}^{q_j q_j \rightarrow q_j q_j}
               \end{array} \right] \, ,
\eeq
with
\beqa
H_{11}^{q_j q_j \rightarrow q_j q_j}
&=&\frac{2}{N_c^2} \left[\frac{(s^2+u^2)}{t^2}
+\frac{1}{N_c^2}\frac{(s^2+t^2)}{u^2}-\frac{2}{N_c}\frac{s^2}{tu}\right] \, ,
\nonumber \\
H_{12}^{q_j q_j \rightarrow q_j q_j}
&=&\frac{2C_F}{N_c^4} \left[N_c\frac{s^2}{tu}
-\frac{(s^2+t^2)}{u^2}\right] \, ,
\nonumber \\
H_{22}^{q_j q_j \rightarrow q_j q_j}
&=&\frac{2C_F^2}{N_c^4} \frac{(s^2+t^2)}{u^2} \, .
\eeqa

Then, the Born cross section is
\beq
E_J \frac{d^3{\hat \sigma}^B_{q_j q_j \rightarrow q_j q_j}}{d^3p_J}
\equiv \sigma^B_{q_j q_j \rightarrow q_j q_j} \delta(s_4)
=\alpha_s^2 \frac{(N_c^2-1)}{2N_c^2\, s} \left[\frac{s^2+u^2}{t^2}
+\frac{s^2+t^2}{u^2}-\frac{2}{N_c}\frac{s^2}{tu}\right] \delta(s_4) \, .
\eeq

The NLO corrections are
\beqa
E_J\frac{d^3{\hat \sigma}^{(1)}_{q_j q_j \rightarrow q_j q_j}}{d^3p_J}&=&
\frac{\alpha_s}{\pi}
\sigma^B_{q_j q_j \rightarrow q_j q_j}
\left\{2C_F \left[\frac{\ln(s_4/p_T^2)}{s_4}\right]_+ \right.
\nonumber \\ &&  \hspace{-20mm} \left.
{}+\left[-2C_F\ln\left(\frac{\mu_F^2}{p_T^2}\right)
-\frac{3}{2}C_F-\frac{(N_c^2+1)}{N_c}\ln\left(\frac{-t}{s}\right)
+\frac{(N_c^2-3)}{N_c}\ln\left(\frac{-u}{s}\right)\right]
\left[\frac{1}{s_4}\right]_+ \right\}
\nonumber \\ && \hspace{-25mm}
{}+\frac{\alpha_s^3}{\pi}\left\{\frac{4C_F^2}{N_c}\ln\left(\frac{t}{u}\right)
\frac{(s^2+t^2)}{u^2}+\frac{8C_F^2}{N_c^2}\ln\left(\frac{-u}{s}\right)
\frac{s^2}{tu}\right\} \left[\frac{1}{s_4}\right]_+
\nonumber \\ && \hspace{-25mm}
{}+\frac{\alpha_s}{\pi}
\sigma^B_{q_j q_j \rightarrow q_j q_j} \delta(s_4)
\left\{-C_F\left[\ln\left(\frac{p_T^2}{s}\right)+\frac{3}{2}\right]
\ln\left(\frac{\mu_F^2}{p_T^2}\right)
+\frac{\beta_0}{2}\ln\left(\frac{\mu_R^2}{p_T^2}\right)\right\} \, .
\eeqa

The NNLO corrections are 
\beqa
E_J\frac{d^3{\hat \sigma}^{(2)}_{q_j q_j \rightarrow q_j q_j}}{d^3p_J}&=&
\left(\frac{\alpha_s}{\pi}\right)^2
\sigma^B_{q_j q_j \rightarrow q_j q_j}
\left\{2C_F^2 \left[\frac{\ln^3(s_4/p_T^2)}{s_4}\right]_+ \right.
\nonumber \\ &&  \hspace{-15mm} 
{}+3C_F\left[-2C_F\ln\left(\frac{\mu_F^2}{p_T^2}\right)
-\frac{3}{2}C_F-\frac{(N_c^2+1)}{N_c}\ln\left(\frac{-t}{s}\right) \right.
\nonumber \\ && \quad \left. \left.
{}+\frac{(N_c^2-3)}{N_c}\ln\left(\frac{-u}{s}\right)-\frac{\beta_0}{12}\right]
\left[\frac{\ln^2(s_4/p_T^2)}{s_4}\right]_+ \right\}
\nonumber \\ &&  \hspace{-20mm} 
{}+\frac{\alpha_s^4}{\pi^2} \frac{12C_F^3}{N_c}
\left[\ln\left(\frac{t}{u}\right)
\frac{(s^2+t^2)}{u^2}+\frac{2}{N_c}\ln\left(\frac{-u}{s}\right)
\frac{s^2}{tu}\right] \left[\frac{\ln^2(s_4/p_T^2)}{s_4}\right]_+
\nonumber \\ &&  \hspace{-20mm} 
{}+\left(\frac{\alpha_s}{\pi}\right)^2
\sigma^B_{q_j q_j \rightarrow q_j q_j}
\left\{4C_F\left[-\frac{1}{2}C_F\ln\left(\frac{p_T^2}{s}\right)
+\frac{3}{4}C_F+\frac{(N_c^2+1)}{N_c}\ln\left(\frac{-t}{s}\right)\right.\right.
\nonumber \\ &&  \hspace{-15mm} \left. \left.
{}-\frac{(N_c^2-3)}{N_c}\ln\left(\frac{-u}{s}\right)
+C_F\ln\left(\frac{\mu_F^2}{p_T^2}\right)\right]
\ln\left(\frac{\mu_F^2}{p_T^2}\right)
+\frac{3}{2}C_F\beta_0\ln\left(\frac{\mu_R^2}{p_T^2}\right) \right\}
\left[\frac{\ln(s_4/p_T^2)}{s_4}\right]_+ 
\nonumber \\ &&  \hspace{-20mm}
{}+\frac{\alpha_s^4}{\pi^2}\frac{16C_F^3}{N_c}
\left[-\ln\left(\frac{t}{u}\right)
\frac{(s^2+t^2)}{u^2}-\frac{2}{N_c}\ln\left(\frac{-u}{s}\right)
\frac{s^2}{tu}\right]\ln\left(\frac{\mu_F^2}{p_T^2}\right) 
\left[\frac{\ln(s_4/p_T^2)}{s_4}\right]_+
\nonumber \\ &&  \hspace{-20mm} 
{}+\left(\frac{\alpha_s}{\pi}\right)^2
\sigma^B_{q_j q_j \rightarrow q_j q_j} \left\{
C_F\left[2C_F\ln\left(\frac{p_T^2}{s}\right)+3C_F+\frac{\beta_0}{4}\right]
\ln^2\left(\frac{\mu_F^2}{p_T^2}\right) \right.
\nonumber \\ && \hspace{15mm} \left.
{}-\frac{3}{2}C_F\beta_0\ln\left(\frac{\mu_F^2}{p_T^2}\right) 
\ln\left(\frac{\mu_R^2}{p_T^2}\right)\right\}\left[\frac{1}{s_4}\right]_+ \, .
\eeqa

\subsection{NNLO corrections for $q_j q_k \rightarrow q_j q_k $}

The hard matrix at lowest order is
\beq
H^{q_j q_k \rightarrow q_j q_k}=\alpha_s^2 \left[
                \begin{array}{cc}
                 2(s^2+u^2)/(N_c^2 t^2) & 0  
\vspace{2mm} \\
                 0 & 0
               \end{array} \right] \, .
\eeq

The Born cross section is then given by
\beq
E_J \frac{d^3{\hat \sigma}^B_{q_j q_k \rightarrow q_j q_k}}{d^3p_J}
\equiv \sigma^B_{q_j q_k \rightarrow q_j q_k} \delta(s_4)
=\alpha_s^2 \frac{(N_c^2-1)}{2N_c^2\, s} \frac{(s^2+u^2)}{t^2} \delta(s_4) \, .
\eeq

The NLO corrections are
\beqa
E_J\frac{d^3{\hat \sigma}^{(1)}_{q_j q_k \rightarrow q_j q_k}}{d^3p_J}&=&
\frac{\alpha_s}{\pi}
\sigma^B_{q_j q_k \rightarrow q_j q_k}
\left\{2C_F \left[\frac{\ln(s_4/p_T^2)}{s_4}\right]_+ \right.
\nonumber \\ &&  \hspace{-20mm} \left.
{}+\left[-2C_F\ln\left(\frac{\mu_F^2}{p_T^2}\right)
-\frac{3}{2}C_F-\frac{(N_c^2+1)}{N_c}\ln\left(\frac{-t}{s}\right)
+\frac{(N_c^2-3)}{N_c}\ln\left(\frac{-u}{s}\right)\right]
\left[\frac{1}{s_4}\right]_+ \right.
\nonumber \\ && \hspace{-20mm} \left.
{}+\delta(s_4)
\left[-C_F\left(\ln\left(\frac{p_T^2}{s}\right)+\frac{3}{2}\right)
\ln\left(\frac{\mu_F^2}{p_T^2}\right)
+\frac{\beta_0}{2}\ln\left(\frac{\mu_R^2}{p_T^2}\right)\right]\right\} \, .
\eeqa

The NNLO corrections are
\beqa
E_J\frac{d^3{\hat \sigma}^{(2)}_{q_j q_k \rightarrow q_j q_k}}{d^3p_J}&=&
\left(\frac{\alpha_s}{\pi}\right)^2
\sigma^B_{q_j q_k \rightarrow q_j q_k}
\left\{2C_F^2 \left[\frac{\ln^3(s_4/p_T^2)}{s_4}\right]_+ \right.
\nonumber \\ &&  \hspace{-20mm} 
{}+3C_F\left[-2C_F\ln\left(\frac{\mu_F^2}{p_T^2}\right)
-\frac{3}{2}C_F-\frac{(N_c^2+1)}{N_c}\ln\left(\frac{-t}{s}\right) \right.
\nonumber \\ && \left. 
+\frac{(N_c^2-3)}{N_c}\ln\left(\frac{-u}{s}\right)-\frac{\beta_0}{12}\right]
\left[\frac{\ln^2(s_4/p_T^2)}{s_4}\right]_+ 
\nonumber \\ &&  \hspace{-20mm} 
{}+4C_F\left[-\frac{1}{2}C_F\ln\left(\frac{p_T^2}{s}\right)
+\frac{3}{4}C_F+\frac{(N_c^2+1)}{N_c}\ln\left(\frac{-t}{s}\right)\right.
\nonumber \\ && \left.
{}-\frac{(N_c^2-3)}{N_c}\ln\left(\frac{-u}{s}\right)
+C_F\ln\left(\frac{\mu_F^2}{p_T^2}\right)\right]
\ln\left(\frac{\mu_F^2}{p_T^2}\right)
\left[\frac{\ln(s_4/p_T^2)}{s_4}\right]_+
\nonumber \\ && \hspace{-20mm}
{}+\frac{3}{2}C_F\beta_0\ln\left(\frac{\mu_R^2}{p_T^2}\right)
\left[\frac{\ln(s_4/p_T^2)}{s_4}\right]_+
\nonumber \\ &&  \hspace{-20mm} 
{}+C_F\left[2C_F\ln\left(\frac{p_T^2}{s}\right)+3C_F+\frac{\beta_0}{4}\right]
\ln^2\left(\frac{\mu_F^2}{p_T^2}\right) \left[\frac{1}{s_4}\right]_+ 
\nonumber \\ &&  \hspace{-20mm} \left.
{}-\frac{3}{2}C_F\beta_0\ln\left(\frac{\mu_F^2}{p_T^2}\right) 
\ln\left(\frac{\mu_R^2}{p_T^2}\right)\left[\frac{1}{s_4}\right]_+ \right\} \, .
\eeqa

\mysection{NNLO-NLL corrections for $q {\bar q} \rightarrow  gg $
and $gg \rightarrow q {\bar q}$}

Next, we analyze the processes $q {\bar q} \rightarrow  gg $ and
$gg \rightarrow q {\bar q}$. 

For the process 
\beq
q\left(p_a, r_a \right)+\bar{q}\left(p_b, r_b \right) \rightarrow
g\left(p_1, r_1 \right)+g\left(p_2, r_2 \right) \, ,
\eeq
in the $s$-channel color basis
\beq
c_1^{q {\bar q} \rightarrow gg}=\delta_{r_a r_b}\delta_{r_1 r_2} \, , \quad
c_2^{q {\bar q} \rightarrow gg}
=d^{r_1 r_2 c}{\left( T_F^c \right)}_{r_b r_a} \, , \quad
c_3^{q {\bar q} \rightarrow gg}
=if^{r_1 r_2 c}{\left( T_F^c \right)}_{r_b r_a} \, ,
\eeq
where $d^{abc}$ and $f^{abc}$ are the totally symmetric and antisymmetric
$SU(3)$ invariant tensors, the soft matrix at lowest order is
\beq
S^{q {\bar q} \rightarrow gg}=\left[
                \begin{array}{ccc}
                 N_c(N_c^2-1) & 0 & 0 
\vspace{2mm} \\
                 0 & (N_c^2-4)(N_c^2-1)/(2N_c) & 0
\vspace{2mm} \\
                 0 & 0 & N_c(N_c^2-1)/2
               \end{array} \right] \, ,
\eeq
and the one-loop soft anomalous dimension matrix is~\cite{Kidonakis:1998nf} 
\beq
\Gamma_{S'}^{q {\bar q} \rightarrow gg}=\frac{\alpha_s}{\pi}\left[
                \begin{array}{ccc}
                 0  &   0  & U-T  \vspace{2mm} \\ 
                 0  &   \frac{C_A}{2}\left(T+U \right)    & \frac{C_A}{2}
\left(U-T\right) \vspace{2mm} \\ 
                 2\left(U-T \right)  & \frac{N_c^2-4}{2N_c}
\left(U-T\right)  & \frac{C_A}{2}\left(T+U \right)
                \end{array} \right].
\label{Gammaqqgg}
\eeq

These same two matrices also describe the time-reversed process
\cite{Kidonakis:1998nf,Kidonakis:1997gm}
\beq
g\left(p_1, r_1 \right)+g\left(p_2, r_2 \right) \rightarrow
{\bar q}\left(p_a, r_a \right)+q\left(p_b, r_b \right) \, .
\eeq
 
\subsection{NNLO corrections for $q {\bar q} \rightarrow  gg$}

For this process the hard matrix at lowest order is given by
\beq
H^{q {\bar q} \rightarrow gg}=\alpha_s^2 \left[
                \begin{array}{ccc}
                 H_{11}^{q {\bar q} \rightarrow gg}
 & H_{12}^{q {\bar q} \rightarrow gg} & H_{13}^{q {\bar q} \rightarrow gg} 
\vspace{2mm} \\
                 H_{12}^{q {\bar q} \rightarrow gg} 
 & H_{22}^{q {\bar q} \rightarrow gg} & H_{23}^{q {\bar q} \rightarrow gg}
\vspace{2mm} \\
                 H_{13}^{q {\bar q} \rightarrow gg}
 & H_{23}^{q {\bar q} \rightarrow gg} & H_{33}^{q {\bar q} \rightarrow gg}
               \end{array} \right] \, ,
\eeq
with
\beqa
H_{11}^{q {\bar q} \rightarrow gg}
&=&\frac{1}{2 N_c^4} \left(\frac{u}{t}+\frac{t}{u}\right) \, ,
\nonumber \\
H_{12}^{q {\bar q} \rightarrow gg}&=& N_c H_{11}^{q {\bar q}\rightarrow gg}\, ,
\nonumber \\
H_{22}^{q {\bar q}\rightarrow gg}&=&N_c^2 H_{11}^{q {\bar q}\rightarrow gg}\,, 
\nonumber \\
H_{13}^{q {\bar q} \rightarrow gg}&=& -\frac{1}{2 N_c^3} \frac{(u^2-t^2)}{tu}
          -\frac{1}{N_c^3} \frac{(u-t)}{s} \, ,
\nonumber \\
H_{23}^{q {\bar q} \rightarrow gg}&=& N_c H_{13}^{q {\bar q}\rightarrow gg}\, ,
\nonumber \\
H_{33}^{q {\bar q} \rightarrow gg}
&=&\frac{1}{2 N_c^2} \frac{s^2}{tu}+\frac{4}{N_c^2} \frac{tu}{s^2}
           -\frac{3}{N_c^2} \, .
\eeqa

The Born cross section is given by
\beq
E_J \frac{d^3{\hat \sigma}^B_{q {\bar q} \rightarrow gg}}{d^3p_J}
\equiv \sigma^B_{q {\bar q} \rightarrow gg} \delta(s_4)
=\alpha_s^2 \frac{(N_c^2-1)}{N_c \, s} \left[\frac{(N_c^2-1)}{2N_c^2} 
\frac{(t^2+u^2)}{tu}-1+2\frac{tu}{s^2}\right] \delta(s_4) \, .
\eeq

The NLO corrections are
\beqa
E_J\frac{d^3{\hat \sigma}^{(1)}_{q {\bar q} \rightarrow gg}}{d^3p_J}
&=&\frac{\alpha_s}{\pi}
\sigma^B_{q {\bar q} \rightarrow gg}
\left\{(4C_F-2C_A) \left[\frac{\ln(s_4/p_T^2)}{s_4}\right]_+ \right.
\nonumber \\ &&  \hspace{-10mm} \left.
{}+\left[-2C_F\ln\left(\frac{\mu_F^2}{p_T^2}\right)
-C_A\ln\left(\frac{p_T^2}{s}\right)-\frac{\beta_0}{2}\right]
\left[\frac{1}{s_4}\right]_+ \right\}
\nonumber \\ && \hspace{-25mm}
{}+\frac{\alpha_s^3}{\pi}\left\{-\frac{(N_c^2-1)}{2N_c^2}
\frac{(t^2+u^2)}{tu}\ln\left(\frac{p_T^2}{s}\right)
-\frac{(N_c^2-1)}{2}\left(\frac{(u^2-t^2)}{tu}+\frac{2(u-t)}{s}\right)
\ln\left(\frac{u}{t}\right)\right\} \left[\frac{1}{s_4}\right]_+
\nonumber \\ && \hspace{-25mm}
{}+\frac{\alpha_s}{\pi}
\sigma^B_{q {\bar q} \rightarrow gg} \delta(s_4)
\left\{-C_F\left[\ln\left(\frac{p_T^2}{s}\right)+\frac{3}{2}\right]
\ln\left(\frac{\mu_F^2}{p_T^2}\right)
+\frac{\beta_0}{2}\ln\left(\frac{\mu_R^2}{p_T^2}\right)\right\} \, .
\eeqa

The NNLO corrections are
\beqa
E_J\frac{d^3{\hat \sigma}^{(2)}_{q {\bar q} \rightarrow gg}}{d^3p_J}&=&
\left(\frac{\alpha_s}{\pi}\right)^2
\sigma^B_{q {\bar q} \rightarrow gg}
\left\{\frac{1}{2}(4C_F-2C_A)^2 \left[\frac{\ln^3(s_4/p_T^2)}{s_4}\right]_+ 
\right.
\nonumber \\ &&  \hspace{-13mm}
{}+\left[3(2C_F-C_A)\left(-2C_F\ln\left(\frac{\mu_F^2}{p_T^2}\right)
-C_A\ln\left(\frac{p_T^2}{s}\right)\right)\right.
\nonumber \\ && \left. \left.
{}+\beta_0 \left(-4C_F+\frac{9}{4}C_A\right)\right]
\left[\frac{\ln^2(s_4/p_T^2)}{s_4}\right]_+ \right\}
\nonumber \\ &&  \hspace{-20mm} 
{}+\frac{\alpha_s^4}{\pi^2} \;  3 \, (2C_F-C_A)
\left[-\frac{(N_c^2-1)}{2N_c^2}
\frac{(t^2+u^2)}{tu}\ln\left(\frac{p_T^2}{s}\right) \right.
\nonumber \\ && \left.
{}-\frac{(N_c^2-1)}{2}\left(\frac{(u^2-t^2)}{tu}+\frac{2(u-t)}{s}\right)
\ln\left(\frac{u}{t}\right)\right]
\left[\frac{\ln^2(s_4/p_T^2)}{s_4}\right]_+
\nonumber \\ &&  \hspace{-20mm} 
{}+\left(\frac{\alpha_s}{\pi}\right)^2
\sigma^B_{q {\bar q} \rightarrow gg}
\left\{4C_F\left[-\left(C_F-\frac{C_A}{2}\right)
\left(\ln\left(\frac{p_T^2}{s}\right)+\frac{3}{2}\right)
+C_A\ln\left(\frac{p_T^2}{s}\right) \right. \right.
\nonumber \\ && \hspace{-15mm} \left. \left.
{}+\frac{\beta_0}{2}
+C_F\ln\left(\frac{\mu_F^2}{p_T^2}\right)\right]
\ln\left(\frac{\mu_F^2}{p_T^2}\right) 
+3\beta_0\left(C_F-\frac{C_A}{2}\right)\ln\left(\frac{\mu_R^2}{p_T^2}\right)
\right\}\left[\frac{\ln(s_4/p_T^2)}{s_4}\right]_+
\nonumber \\ &&  \hspace{-20mm}
{}+\frac{\alpha_s^4}{\pi^2} \; 4 C_F \ln\left(\frac{\mu_F^2}{p_T^2}\right)
\left[\frac{(N_c^2-1)}{2N_c^2}
\frac{(t^2+u^2)}{tu}\ln\left(\frac{p_T^2}{s}\right) \right.
\nonumber \\ && \left.
{}+\frac{(N_c^2-1)}{2}\left(\frac{(u^2-t^2)}{tu}+\frac{2(u-t)}{s}\right)
\ln\left(\frac{u}{t}\right)\right]
\left[\frac{\ln(s_4/p_T^2)}{s_4}\right]_+
\nonumber \\ &&  \hspace{-20mm} 
{}+\left(\frac{\alpha_s}{\pi}\right)^2
\sigma^B_{q {\bar q} \rightarrow gg} \left\{
C_F\left[2C_F\ln\left(\frac{p_T^2}{s}\right)+3C_F+\frac{\beta_0}{4}\right]
\ln^2\left(\frac{\mu_F^2}{p_T^2}\right) \right.
\nonumber \\ &&  \hspace{15mm} \left.
{}-\frac{3}{2}C_F\beta_0\ln\left(\frac{\mu_F^2}{p_T^2}\right) 
\ln\left(\frac{\mu_R^2}{p_T^2}\right)\right\}\left[\frac{1}{s_4}\right]_+ \, .
\eeqa

\subsection{NNLO corrections for $gg \rightarrow  q {\bar q} $}

Note that apart from color factors, the hard matrix for this process
is the same as for $q {\bar q} \rightarrow gg$:
\beq
H^{gg \rightarrow q {\bar q}}=\frac{N_c^2}{(N_c^2-1)^2} 
H^{q {\bar q} \rightarrow gg} \, .
\eeq

Then for the process 
\beq
g\left(p_a, r_a \right)+g\left(p_b, r_b \right) \rightarrow
q\left(p_1, r_1 \right)+{\bar q}\left(p_2, r_2 \right) \, ,
\eeq
the Born cross section (symmetric in $t$,$u$) is given by
\beq
E_J \frac{d^3{\hat \sigma}^B_{gg \rightarrow q {\bar q}}}{d^3p_J}
\equiv \sigma^B_{gg \rightarrow q {\bar q}} \delta(s_4)
=\alpha_s^2 \frac{N_c}{(N_c^2-1)\, s} \left[\frac{(N_c^2-1)}{2N_c^2}
\frac{(t^2+u^2)}{tu}-1+2\frac{tu}{s^2}\right]  \delta(s_4) \, .
\eeq

The NLO corrections are
\beqa
E_J\frac{d^3{\hat \sigma}^{(1)}_{gg \rightarrow q {\bar q}}}{d^3p_J}
&=&\frac{\alpha_s}{\pi}
\sigma^B_{gg \rightarrow q {\bar q}}
\left\{(4C_A-2C_F) \left[\frac{\ln(s_4/p_T^2)}{s_4}\right]_+ \right.
\nonumber \\ &&  \hspace{-10mm} \left.
{}+\left[-\frac{3}{2}C_F-(2C_F-C_A)\ln\left(\frac{p_T^2}{s}\right)
-2C_A\ln\left(\frac{\mu_F^2}{p_T^2}\right)\right]
\left[\frac{1}{s_4}\right]_+ \right\}
\nonumber \\ && \hspace{-20mm}
{}+\frac{\alpha_s^3}{\pi}\left\{-\frac{(N_c^2-1)}{2N_c^2}
\frac{(t^2+u^2)}{tu}\ln\left(\frac{p_T^2}{s}\right)
-\frac{(N_c^2-1)}{2}\left(\frac{(u^2-t^2)}{tu}+\frac{2(u-t)}{s}\right)
\ln\left(\frac{u}{t}\right)\right\} \left[\frac{1}{s_4}\right]_+
\nonumber \\ && \hspace{-20mm}
{}+\frac{\alpha_s}{\pi}
\sigma^B_{gg \rightarrow q {\bar q}} \delta(s_4)
\left\{-C_A \ln\left(\frac{p_T^2}{s}\right)
\ln\left(\frac{\mu_F^2}{p_T^2}\right)
+\frac{\beta_0}{2}\ln\left(\frac{\mu_R^2}{\mu_F^2}\right)\right\} \, .
\eeqa

The NNLO corrections are
\beqa
E_J\frac{d^3{\hat \sigma}^{(2)}_{gg \rightarrow q {\bar q}}}{d^3p_J}&=&
\left(\frac{\alpha_s}{\pi}\right)^2
\sigma^B_{gg \rightarrow q {\bar q}}
\left\{\frac{1}{2}(4C_A-2C_F)^2 \left[\frac{\ln^3(s_4/p_T^2)}{s_4}\right]_+ 
\right.
\nonumber \\ &&  \hspace{-13mm}
{}+\left[3(2C_A-C_F)\left(-\frac{3}{2}C_F
-(2C_F-C_A)\ln\left(\frac{p_T^2}{s}\right)
-2C_A\ln\left(\frac{\mu_F^2}{p_T^2}\right)\right) \right.
\nonumber \\ && \left. \left.
{}+\beta_0 \left(-C_A+\frac{3}{4}C_F\right)\right]
\left[\frac{\ln^2(s_4/p_T^2)}{s_4}\right]_+ \right\}
\nonumber \\ &&  \hspace{-20mm} 
{}+\frac{\alpha_s^4}{\pi^2} \;  3 \, (2C_A-C_F)
\left[-\frac{(N_c^2-1)}{2N_c^2}
\frac{(t^2+u^2)}{tu}\ln\left(\frac{p_T^2}{s}\right) \right.
\nonumber \\ && \left.
{}-\frac{(N_c^2-1)}{2}\left(\frac{(u^2-t^2)}{tu}+\frac{2(u-t)}{s}\right)
\ln\left(\frac{u}{t}\right)\right]
\left[\frac{\ln^2(s_4/p_T^2)}{s_4}\right]_+
\nonumber \\ &&  \hspace{-20mm} 
{}+\left(\frac{\alpha_s}{\pi}\right)^2
\sigma^B_{gg \rightarrow q {\bar q}}
\left\{4C_A\left[\left(\frac{5}{2}C_F-2C_A\right)
\ln\left(\frac{p_T^2}{s}\right)+\frac{3}{2}C_F
+C_A\ln\left(\frac{\mu_F^2}{p_T^2}\right)\right] 
\ln\left(\frac{\mu_F^2}{p_T^2}\right)
\right. 
\nonumber \\ && \hspace{-15mm} \left. 
{}-\beta_0 (2C_A-C_F)\ln\left(\frac{\mu_F^2}{p_T^2}\right)
+\frac{3}{2} \beta_0 (2C_A-C_F)
\ln\left(\frac{\mu_R^2}{p_T^2}\right)\right\}
\left[\frac{\ln(s_4/p_T^2)}{s_4}\right]_+
\nonumber \\ &&  \hspace{-20mm}
{}+\frac{\alpha_s^4}{\pi^2} \; 4 C_A \ln\left(\frac{\mu_F^2}{p_T^2}\right)
\left[\frac{(N_c^2-1)}{2N_c^2}
\frac{(t^2+u^2)}{tu}\ln\left(\frac{p_T^2}{s}\right) \right.
\nonumber \\ && \left.
{}+\frac{(N_c^2-1)}{2}\left(\frac{(u^2-t^2)}{tu}+\frac{2(u-t)}{s}\right)
\ln\left(\frac{u}{t}\right)\right]
\left[\frac{\ln(s_4/p_T^2)}{s_4}\right]_+
\nonumber \\ &&  \hspace{-20mm} 
{}+\left(\frac{\alpha_s}{\pi}\right)^2
\sigma^B_{gg \rightarrow q {\bar q}} \left\{
C_A\left[2C_A\ln\left(\frac{p_T^2}{s}\right)+\frac{5}{4}\beta_0\right]
\ln^2\left(\frac{\mu_F^2}{p_T^2}\right) \right.
\nonumber \\ &&  \hspace{15mm} \left.
{}-\frac{3}{2}C_A\beta_0\ln\left(\frac{\mu_F^2}{p_T^2}\right)
\ln\left(\frac{\mu_R^2}{p_T^2}\right)\right\}\left[\frac{1}{s_4}\right]_+ \, .
\eeqa

\mysection{NNLO-NLL corrections for $q g \rightarrow  q g$}

Here we discuss quark-gluon scattering,
\beq
q\left(p_a, r_a \right)+g\left(p_b, r_b \right) \rightarrow
q\left(p_1, r_1 \right)+g\left(p_2, r_2 \right) \, .
\eeq
In the $t$-channel color basis
\beq
c_1^{q g \rightarrow  q g}=\delta_{r_a r_1}\delta_{r_b r_2} \, , \quad
c_2^{q g \rightarrow  q g}
=d^{r_b r_2 c}{\left( T_F^c \right)}_{r_1 r_a} \, , \quad
c_3^{q g \rightarrow  q g}=if^{r_b r_2 c}{\left( T_F^c \right)}_{r_1 r_a} \, ,
\label{eq:basqgqg}
\eeq
the soft matrix at lowest order is
\beq
S^{q g \rightarrow  q g}=\left[
                \begin{array}{ccc}
                 N_c(N_c^2-1) & 0 & 0 
\vspace{2mm} \\
                 0 & (N_c^2-4)(N_c^2-1)/(2N_c) & 0
\vspace{2mm} \\
                 0 & 0 & N_c(N_c^2-1)/2
               \end{array} \right]\, ,
\eeq
and the one-loop soft anomalous dimension matrix is~\cite{Kidonakis:1998nf}
\beq
\Gamma_{S'}^{q g \rightarrow  q g}=\frac{\alpha_s}{\pi}\left[
                \begin{array}{ccc}
                 \left( C_F+C_A \right) T  &   0  & U  \vspace{2mm} \\ 
                 0  &   C_F T+ \frac{C_A}{2} U     & \frac{C_A}{2} U  
\vspace{2mm} \\
                 2 U  & \frac{N_c^2-4}{2N_c} U  &  C_F T+ \frac{C_A}{2} U
                \end{array} \right] \, .
\label{Gammaqgqg}
\eeq

The hard matrix at lowest order is given by 
(see also Ref.~\cite{Oderda:2000kr})
\beq
H^{q g \rightarrow  q g}=\alpha_s^2 \left[
                \begin{array}{ccc}
                 H_{11}^{q g \rightarrow  q g} & H_{12}^{q g \rightarrow  q g}
 & H_{13}^{q g \rightarrow  q g} 
\vspace{2mm} \\
                 H_{12}^{q g \rightarrow  q g} & H_{22}^{q g \rightarrow  q g}
 & H_{23}^{q g \rightarrow  q g}
\vspace{2mm} \\
                 H_{13}^{q g \rightarrow  q g} & H_{23}^{q g \rightarrow  q g}
 & H_{33}^{q g \rightarrow  q g}
               \end{array} \right] \, ,
\eeq
with
\beqa
H_{11}^{q g \rightarrow  q g}
&=&-\frac{1}{2 N_c^3 (N_c^2-1)} \left(\frac{t^2}{su}-2\right) \, ,
\nonumber \\
H_{12}^{q g \rightarrow  q g}&=& N_c H_{11}^{q g \rightarrow  q g} \, ,
\nonumber \\
H_{22}^{q g \rightarrow  q g}&=& N_c^2 H_{11}^{q g \rightarrow  q g} \, ,
\nonumber \\
H_{13}^{q g \rightarrow  q g}
&=& \frac{1}{N_c^2 (N_c^2-1)} \left[-1-\frac{2s}{t}
          +\frac{u}{2s}-\frac{s}{2u}\right] \, ,
\nonumber \\
H_{23}^{q g \rightarrow  q g}&=& N_c H_{13}^{q g \rightarrow  q g} \, ,
\nonumber \\
H_{33}^{q g \rightarrow  q g}
&=&\frac{1}{N_c (N_c^2-1)} \left[3-\frac{4su}{t^2}
         -\frac{t^2}{2su} \right] \, .
\eeqa

The Born cross section is
\beq
E_J \frac{d^3{\hat \sigma}^B_{q g \rightarrow q g}}{d^3p_J}
\equiv \sigma^B_{q g \rightarrow q g} \delta(s_4)
=\alpha_s^2 \frac{1}{s} \left[2-\frac{1}{N_c^2}-\frac{(N_c^2-1)}{2N_c^2}
\frac{t^2}{su}-2\frac{su}{t^2}\right] \delta(s_4) \, .
\eeq

The NLO corrections are
\beqa
E_J\frac{d^3{\hat \sigma}^{(1)}_{q g \rightarrow q g}}{d^3p_J}
&=&\frac{\alpha_s}{\pi}
\sigma^B_{q g \rightarrow q g}
\left\{(C_F+C_A) \left[\frac{\ln(s_4/p_T^2)}{s_4}\right]_+ \right.
\nonumber \\ &&  \hspace{-13mm} \left.
{}+\left[-(C_F+C_A)\ln\left(\frac{\mu_F^2}{p_T^2}\right)
+C_F\left(-2\ln\left(\frac{-u}{s}\right)-\frac{3}{4}\right)
-2C_A\ln\left(\frac{-t}{s}\right)-\frac{\beta_0}{4}\right]
\left[\frac{1}{s_4}\right]_+ \right\}
\nonumber \\ && \hspace{-25mm}
{}+\frac{\alpha_s^3}{\pi}\left\{-\frac{1}{N_c^2} (C_F+C_A)
\left(\frac{t^2}{su}-2\right)\ln\left(\frac{-t}{s}\right)
+N_c\left(-1-\frac{2s}{t}+\frac{u}{2s}-\frac{s}{2u}\right)
\ln\left(\frac{-u}{s}\right) \right.
\nonumber \\ && \hspace{-13mm} \left.
{}+\left(C_F\ln\left(\frac{-t}{s}\right)
+\frac{C_A}{2}\ln\left(\frac{-u}{s}\right)\right)
\left[\left(\frac{2}{N_c^2}-1\right)\left(\frac{t^2}{su}-2\right)
+2\left(1-2\frac{su}{t^2}\right)\right]\right\} \left[\frac{1}{s_4}\right]_+
\nonumber \\ && \hspace{-25mm}
{}+\frac{\alpha_s}{\pi}
\sigma^B_{q g \rightarrow q g} \delta(s_4)
\left\{-\left[C_F\ln\left(\frac{-t}{s}\right)
+C_A\ln\left(\frac{-u}{s}\right)+\frac{3}{4}C_F+\frac{\beta_0}{4}\right]
\ln\left(\frac{\mu_F^2}{p_T^2}\right)
+\frac{\beta_0}{2}\ln\left(\frac{\mu_R^2}{p_T^2}\right)\right\} \, .
\nonumber \\
\eeqa

The NNLO corrections are
\beqa
E_J\frac{d^3{\hat \sigma}^{(2)}_{q g \rightarrow q g}}{d^3p_J}&=&
\left(\frac{\alpha_s}{\pi}\right)^2
\sigma^B_{q g \rightarrow q g}
\left\{\frac{1}{2}(C_F+C_A)^2 \left[\frac{\ln^3(s_4/p_T^2)}{s_4}\right]_+ 
\right.
\nonumber \\ &&  \quad
{}+\frac{3}{2}(C_F+C_A)\left[-(C_F+C_A)\ln\left(\frac{\mu_F^2}{p_T^2}\right)
+C_F\left(-2\ln\left(\frac{-u}{s}\right)-\frac{3}{4}\right)\right.
\nonumber \\ && \hspace{35mm} \left. \left.
{}-2C_A\ln\left(\frac{-t}{s}\right)
-\frac{\beta_0}{3}\right]
\left[\frac{\ln^2(s_4/p_T^2)}{s_4}\right]_+ \right\} 
\nonumber \\ &&  \hspace{-35mm} 
{}+\frac{\alpha_s^4}{\pi^2} \;  \frac{3}{2} \, (C_F+C_A)
\left\{-\frac{1}{N_c^2} (C_F+C_A)
\left(\frac{t^2}{su}-2\right)\ln\left(\frac{-t}{s}\right) \right.
+N_c\left(-1-\frac{2s}{t}+\frac{u}{2s}-\frac{s}{2u}\right)
\ln\left(\frac{-u}{s}\right)
\nonumber \\ && \hspace{-28mm} \left.
{}+\left(C_F\ln\left(\frac{-t}{s}\right)
+\frac{C_A}{2}\ln\left(\frac{-u}{s}\right)\right)
\left[\left(\frac{2}{N_c^2}-1\right)\left(\frac{t^2}{su}-2\right)
+2\left(1-2\frac{su}{t^2}\right)\right]\right\}
\left[\frac{\ln^2(s_4/p_T^2)}{s_4}\right]_+
\nonumber \\ &&  \hspace{-35mm} 
{}+\left(\frac{\alpha_s}{\pi}\right)^2
\sigma^B_{q g \rightarrow q g}
(C_F+C_A)\left\{\left[(4C_F-C_A)
\ln\left(\frac{-u}{s}\right)
+(4C_A-C_F)\ln\left(\frac{-t}{s}\right)
+\frac{3}{4}C_F+\frac{\beta_0}{4} \right. \right.
\nonumber \\ && \hspace{-25mm} \left. \left. 
{}+(C_F+C_A)\ln\left(\frac{\mu_F^2}{p_T^2}\right)\right]
\ln\left(\frac{\mu_F^2}{p_T^2}\right) 
+\frac{3}{4}\beta_0 \ln\left(\frac{\mu_R^2}{p_T^2}\right)
\right\}\left[\frac{\ln(s_4/p_T^2)}{s_4}\right]_+
\nonumber \\ &&  \hspace{-35mm}
{}+\frac{\alpha_s^4}{\pi^2} \; (-2) (C_F+C_A) 
\left\{-\frac{1}{N_c^2}(C_F+C_A)
\left(\frac{t^2}{su}-2\right)\ln\left(\frac{-t}{s}\right)
+N_c\left(-1-\frac{2s}{t}+\frac{u}{2s}-\frac{s}{2u}\right)
\ln\left(\frac{-u}{s}\right) \right.
\nonumber \\ && \hspace{-34mm} \left.
{}+\left(C_F\ln\left(\frac{-t}{s}\right)
+\frac{C_A}{2}\ln\left(\frac{-u}{s}\right)\right)
\left[\left(\frac{2}{N_c^2}-1\right)\left(\frac{t^2}{su}-2\right)
+2\left(1-2\frac{su}{t^2}\right)\right]\right\}
\ln\left(\frac{\mu_F^2}{p_T^2}\right)
\left[\frac{\ln(s_4/p_T^2)}{s_4}\right]_+
\nonumber \\ &&  \hspace{-35mm} 
{}+\left(\frac{\alpha_s}{\pi}\right)^2
\sigma^B_{q g \rightarrow q g} (C_F+C_A)\left\{\left[
C_F\ln\left(\frac{-t}{s}\right)
+C_A\ln\left(\frac{-u}{s}\right)
+\frac{3}{4}C_F+\frac{3}{8}\beta_0\right]
\ln^2\left(\frac{\mu_F^2}{p_T^2}\right) \right.
\nonumber \\ &&  \hspace{20mm} \left. 
{}-\frac{3}{4}\beta_0\ln\left(\frac{\mu_F^2}{p_T^2}\right) 
\ln\left(\frac{\mu_R^2}{p_T^2}\right)\right\}
\left[\frac{1}{s_4}\right]_+ \, .
\eeqa

\mysection{NNLO-NLL corrections for $g g \rightarrow  g g $}

Finally, we consider gluon-gluon scattering,
\beq
g\left(p_a, r_a \right)+g\left(p_b, r_b \right) \rightarrow
g\left(p_1, r_1 \right)+g\left(p_2, r_2 \right) \, .
\eeq
The color decomposition for this process is by far the most complicated.
For simplicity in this section we use $N_c=3$ explicitly.
A complete color basis for the process $gg \rightarrow gg$
is given by the eight color structures~\cite{Kidonakis:1998nf}
\beqa
c_1^{g g \rightarrow  g g}&=&\frac{i}{4}\left[f^{r_a r_b l}
d^{r_1 r_2 l} - d^{r_a r_b l}f^{r_1 r_2 l}\right] \, , \quad
c_2^{g g \rightarrow  g g}=\frac{i}{4}\left[f^{r_a r_b l}
d^{r_1 r_2 l} + d^{r_a r_b l}f^{r_1 r_2 l}\right] \, ,
\nonumber \\ 
c_3^{g g \rightarrow  g g}&=&\frac{i}{4}\left[f^{r_a r_1 l}
d^{r_b r_2 l}+d^{r_a r_1 l}f^{r_b r_2 l}\right] \, , \quad
c_4^{g g \rightarrow  g g}=\frac{1}{8}\delta_{r_a r_1} 
\delta_{r_b r_2} \, ,
\nonumber \\
c_5^{g g \rightarrow  g g}&=&\frac{3}{5} d^{r_ar_1c} d^{r_br_2c} \, , \quad
c_6^{g g \rightarrow  g g}=\frac{1}{3} f^{r_ar_1c} f^{r_br_2c} \, ,
\nonumber \\
c_7^{g g \rightarrow  g g}&=&    
\frac{1}{2}(\delta_{r_a r_b} \delta_{r_1 r_2}
-\delta_{r_a r_2} \delta_{r_b r_1})
-\frac{1}{3} f^{r_ar_1c} f^{r_br_2c} \, ,
\nonumber \\
c_8^{g g \rightarrow  g g}&=&\frac{1}{2}(\delta_{r_a r_b} 
\delta_{r_1 r_2} +\delta_{r_a r_2} \delta_{r_b r_1})
-\frac{1}{8}\delta_{r_a r_1} \delta_{r_b r_2}
-\frac{3}{5} d^{r_ar_1c} d^{r_br_2c} \, .
\label{8x8basis}
\eeqa

The soft matrix at lowest order is
\beq
S^{g g \rightarrow  g g}=   \left[ 
           \begin{array}{cc}
                 S_{3 \times 3}^{g g \rightarrow  g g} & 0_{3 \times 5}
\vspace{2mm} \\
                 0_{5 \times 3} & S_{5 \times 5}^{g g \rightarrow  g g}
               \end{array} \right] \, ,
\eeq
where
\beq
S_{3 \times 3}^{g g \rightarrow  g g}= \left[
                \begin{array}{ccc}
                 5 & 0 & 0  
\vspace{2mm} \\  0 & 5 & 0 
\vspace{2mm} \\  0 & 0 & 5
               \end{array} \right] \, ,
\quad 
S_{5 \times 5}^{g g \rightarrow  g g}= \left[
                \begin{array}{ccccc}
                 1 & 0 & 0 & 0  & 0
\vspace{2mm} \\  0 & 8 & 0 & 0  & 0
\vspace{2mm} \\  0 & 0 & 8 & 0  & 0
\vspace{2mm} \\  0 & 0 & 0 & 20 & 0
\vspace{2mm} \\  0 & 0 & 0 & 0  & 27
       
               \end{array} \right] \, .
\eeq

The one-loop soft anomalous dimension matrix is~\cite{Kidonakis:1998nf}
\beq
\Gamma_{S'}^{g g \rightarrow  g g}=\left[\begin{array}{cc}
            \Gamma_{3 \times 3}^{g g \rightarrow  g g} & 0_{3 \times 5} \\
              0_{5 \times 3}      & \Gamma_{5 \times 5}^{g g \rightarrow  g g}
\end{array} \right] \, ,
\label{gammagggg}
\eeq
with
\beq
\Gamma_{3 \times 3}^{g g \rightarrow  g g}=\frac{\alpha_s}{\pi} \left[
                \begin{array}{ccc}
                  3 T  &   0  & 0  \\
                  0  &  3 U & 0    \\
                  0  &  0  &  3\left( T+ U \right)
                   \end{array} \right]
\eeq
and
\beq
\Gamma_{5 \times 5}^{g g \rightarrow  g g}
=\frac{\alpha_s}{\pi}\left[\begin{array}{ccccc}
6 T & 0 & -6 U & 0 & 0 \vspace{2mm} \\ 
0  & 3 T+\frac{3 U}{2} & -\frac{3 U}{2} & -3 U & 0 \vspace{2mm} \\ 
-\frac{3 U}{4} & -\frac{3 U}{2} &3 T+\frac{3 U}{2} & 0 & -\frac{9 U}{4} 
\vspace{2mm} \\
0 & -\frac{6 U}{5} & 0 & 3 U & -\frac{9 U}{5} \vspace{2mm} \\
0 & 0 &-\frac{2 U}{3} &-\frac{4 U}{3} & -2 T+4 U
\end{array} \right] \, .
\eeq

The hard matrix at lowest order is \cite{Oderda:2000kr}
\beq
H^{g g \rightarrow  g g}=\alpha_s^2 \left[
                \begin{array}{cc}
                 0_{3 \times 3} & 0_{3 \times 5}
\vspace{2mm} \\
                 0_{5 \times 3} & H_{5 \times 5}^{g g \rightarrow  g g}
               \end{array} \right] \, ,
\eeq
where
\beq
H_{5 \times 5}^{g g \rightarrow  g g}= \left[
                \begin{array}{ccccc}
                 H_{11}^{g g \rightarrow  g g} & H_{12}^{g g \rightarrow  g g}
 & H_{13}^{g g \rightarrow  g g} & 0 & H_{15}^{g g \rightarrow  g g}
\vspace{2mm} \\  H_{12}^{g g \rightarrow  g g} & H_{22}^{g g \rightarrow  g g}
 & H_{23}^{g g \rightarrow  g g} & 0 & H_{25}^{g g \rightarrow  g g}
\vspace{2mm} \\  H_{13}^{g g \rightarrow  g g} & H_{23}^{g g \rightarrow  g g}
 & H_{33}^{g g \rightarrow  g g} & 0 & H_{35}^{g g \rightarrow  g g}
\vspace{2mm} \\     0   &    0   &    0   & 0 & 0
\vspace{2mm} \\  H_{15}^{g g \rightarrow  g g} & H_{25}^{g g \rightarrow  g g}
 & H_{35}^{g g \rightarrow  g g} & 0 & H_{55}^{g g \rightarrow  g g}
       
               \end{array} \right] \, ,
\eeq

with
\beqa
H_{11}^{g g \rightarrow  g g}&=&
\frac{9}{16} \left(1-\frac{tu}{s^2}-\frac{st}{u^2}
+\frac{t^2}{su}\right) \, ,
\nonumber \\
H_{12}^{g g \rightarrow  g g}&=& \frac{1}{2} H_{11}^{g g \rightarrow  g g}\, ,
\nonumber \\
H_{13}^{g g \rightarrow  g g}
&=& \frac{9}{32}\left(\frac{st}{u^2}-\frac{tu}{s^2}
+\frac{u^2}{st}-\frac{s^2}{tu}\right) \, ,
\nonumber \\
H_{15}^{g g \rightarrow  g g}&=&-\frac{1}{3} 
H_{11}^{g g \rightarrow  g g} \, , \quad
H_{22}^{g g \rightarrow  g g}=\frac{1}{4} H_{11}^{g g \rightarrow  g g} \, ,
\nonumber \\
H_{23}^{g g \rightarrow  g g}&=& \frac{1}{2} H_{13}^{g g \rightarrow  g g}
 \, , \quad
H_{25}^{g g \rightarrow  g g}=-\frac{1}{6} H_{11}^{g g \rightarrow  g g}  \, ,
\nonumber \\
H_{33}^{g g \rightarrow  g g}
&=& \frac{27}{64}-\frac{9}{16}\left(\frac{su}{t^2}+\frac{tu}{4s^2}
+\frac{st}{4u^2}\right)+\frac{9}{32}\left(\frac{u^2}{st}
+\frac{s^2}{tu}-\frac{t^2}{2su}\right) \, ,
\nonumber \\
H_{35}^{g g \rightarrow  g g}&=& -\frac{1}{3} H_{13}^{g g \rightarrow  g g}
 \, , \quad
H_{55}^{g g \rightarrow  g g}= \frac{1}{9} H_{11}^{g g \rightarrow  g g} \, .
\eeqa

The Born cross section is
\beq
E_J \frac{d^3{\hat \sigma}^B_{gg \rightarrow gg}}{d^3p_J}
\equiv \sigma^B_{g g \rightarrow g g} \delta(s_4)
=\alpha_s^2 \frac{1}{s}\left[ \frac{27}{2}-\frac{9}{2}\left(\frac{su}{t^2}
    +\frac{tu}{s^2}+\frac{st}{u^2}\right) \right] \delta(s_4) \, .
\eeq

The NLO corrections are
\beqa
E_J\frac{d^3{\hat \sigma}^{(1)}_{g g \rightarrow g g}}{d^3p_J}
&=&\frac{\alpha_s}{\pi}
\sigma^B_{g g \rightarrow g g}
\left\{ 2C_A \left[\frac{\ln(s_4/p_T^2)}{s_4}\right]_+ \right.
\nonumber \\ &&  \hspace{-10mm} \left.
{}+\left[-2C_A \ln\left(\frac{\mu_F^2}{p_T^2}\right)
-2C_A \ln\left(\frac{p_T^2}{s}\right)-\frac{\beta_0}{2}\right]
\left[\frac{1}{s_4}\right]_+ \right\}
\nonumber \\ && \hspace{-30mm}
{}+\frac{\alpha_s^3}{\pi}\left\{\frac{27}{8} 
\left[2\ln\left(\frac{-t}{s}\right)+5\ln\left(\frac{-u}{s}\right)\right]
\left[1-\frac{tu}{s^2}-\frac{st}{u^2}+\frac{t^2}{su}\right]
-\frac{27}{4}\ln\left(\frac{-u}{s}\right)
\left[\frac{st}{u^2}-\frac{tu}{s^2}+\frac{u^2}{st}-\frac{s^2}{tu}\right]
\right.
\nonumber \\ && \hspace{-25mm} \left.
{}+\left[3\ln\left(\frac{-t}{s}\right)
+\frac{3}{2}\ln\left(\frac{-u}{s}\right)\right]
\left[\frac{27}{4}-9\left(\frac{su}{t^2}+\frac{tu}{4s^2}+\frac{st}{4u^2}\right)
+\frac{9}{2}\left(\frac{u^2}{st}+\frac{s^2}{tu}-\frac{t^2}{2su}\right)\right]
\right\} \left[\frac{1}{s_4}\right]_+
\nonumber \\ && \hspace{-30mm}
{}+\frac{\alpha_s}{\pi}
\sigma^B_{g g \rightarrow g g} \delta(s_4)
\left\{-C_A\ln\left(\frac{p_T^2}{s}\right)
\ln\left(\frac{\mu_F^2}{p_T^2}\right)
+\frac{\beta_0}{2}\ln\left(\frac{\mu_R^2}{\mu_F^2}\right) \right\} \, .
\eeqa

The NNLO corrections are
\beqa
E_J\frac{d^3{\hat \sigma}^{(2)}_{g g \rightarrow g g}}{d^3p_J}&=&
\left(\frac{\alpha_s}{\pi}\right)^2
\sigma^B_{g g \rightarrow g g}
\left\{2 C_A^2 \left[\frac{\ln^3(s_4/p_T^2)}{s_4}\right]_+ \right.
\nonumber \\ &&  \hspace{-10mm} \left.
{}+3\, C_A \left[-2 C_A  \ln\left(\frac{\mu_F^2}{p_T^2}\right)
-2 C_A \ln\left(\frac{p_T^2}{s}\right)-\frac{7}{12}\beta_0\right]
\left[\frac{\ln^2(s_4/p_T^2)}{s_4}\right]_+ \right\}
\nonumber \\ &&  \hspace{-35mm} 
{}+\frac{\alpha_s^4}{\pi^2} \; 3 \, C_A
\left\{\frac{27}{8} 
\left[2\ln\left(\frac{-t}{s}\right)+5\ln\left(\frac{-u}{s}\right)\right]
\left[1-\frac{tu}{s^2}-\frac{st}{u^2}+\frac{t^2}{su}\right]
-\frac{27}{4}\ln\left(\frac{-u}{s}\right)
\left[\frac{st}{u^2}-\frac{tu}{s^2}+\frac{u^2}{st}-\frac{s^2}{tu}\right]
\right.
\nonumber \\ && \hspace{-32mm} \left.
{}+\left[3\ln\left(\frac{-t}{s}\right)
+\frac{3}{2}\ln\left(\frac{-u}{s}\right)\right]
\left[\frac{27}{4}-9\left(\frac{su}{t^2}+\frac{tu}{4s^2}+\frac{st}{4u^2}\right)
+\frac{9}{2}\left(\frac{u^2}{st}+\frac{s^2}{tu}-\frac{t^2}{2su}\right)\right]
\right\}
\left[\frac{\ln^2(s_4/p_T^2)}{s_4}\right]_+
\nonumber \\ &&  \hspace{-35mm} 
{}+\left(\frac{\alpha_s}{\pi}\right)^2
\sigma^B_{g g \rightarrow g g}
\left\{C_A\left[ 6 C_A \ln\left(\frac{p_T^2}{s}\right)
+\beta_0+4C_A\ln\left(\frac{\mu_F^2}{p_T^2}\right)\right]
\ln\left(\frac{\mu_F^2}{p_T^2}\right) 
+\frac{3}{2}C_A\beta_0 \ln\left(\frac{\mu_R^2}{p_T^2}\right)
\right\}\left[\frac{\ln(s_4/p_T^2)}{s_4}\right]_+
\nonumber \\ &&  \hspace{-35mm}
{}+\frac{\alpha_s^4}{\pi^2} \; (-4) C_A 
\left\{\frac{27}{8} 
\left[2\ln\left(\frac{-t}{s}\right)+5\ln\left(\frac{-u}{s}\right)\right]
\left[1-\frac{tu}{s^2}-\frac{st}{u^2}+\frac{t^2}{su}\right]
-\frac{27}{4}\ln\left(\frac{-u}{s}\right)
\left[\frac{st}{u^2}-\frac{tu}{s^2}+\frac{u^2}{st}-\frac{s^2}{tu}\right]
\right.
\nonumber \\ && \hspace{-35mm} \left.
{}+\left[3\ln\left(\frac{-t}{s}\right)
+\frac{3}{2}\ln\left(\frac{-u}{s}\right)\right]
\left[\frac{27}{4}-9\left(\frac{su}{t^2}+\frac{tu}{4s^2}+\frac{st}{4u^2}\right)
+\frac{9}{2}\left(\frac{u^2}{st}+\frac{s^2}{tu}-\frac{t^2}{2su}\right)\right]
\right\}
\ln\left(\frac{\mu_F^2}{p_T^2}\right)
\left[\frac{\ln(s_4/p_T^2)}{s_4}\right]_+
\nonumber \\ &&  \hspace{-35mm} 
{}+\left(\frac{\alpha_s}{\pi}\right)^2
\sigma^B_{g g \rightarrow g g} \left\{C_A\left[
2 C_A \ln\left(\frac{p_T^2}{s}\right)
+\frac{5\beta_0}{4} \right]
\ln^2\left(\frac{\mu_F^2}{p_T^2}\right) 
-\frac{3\beta_0}{2}C_A\ln\left(\frac{\mu_F^2}{p_T^2}\right) 
\ln\left(\frac{\mu_R^2}{p_T^2}\right)\right\}
\left[\frac{1}{s_4}\right]_+ \, .
\eeqa

\end{document}